\documentclass[10pt, twocolumn, twoside]{IEEEtran}
 \pdfoutput=1 
\usepackage{amsfonts}
\usepackage{amsmath}
\usepackage{cases}
\usepackage[final]{graphicx}
\usepackage{epsfig}
\usepackage{cite}
\usepackage{float}
\usepackage[bottom]{footmisc}
\usepackage{multirow}
\usepackage{array}
\usepackage{algorithm}
\usepackage{algorithmic}
\usepackage{changebar}

\begin{document}
\newtheorem{corollary}{Corollary}
\newtheorem{lemma}{Lemma}
\newtheorem{proposition}{Proposition}

\title{Interference Alignment for Multicell Multiuser MIMO Uplink Channels}

\author{Khanh~Pham~and~Kyungchun~Lee*,\IEEEmembership{~Member,~IEEE} %
\thanks{This research was supported by Basic Science Research Program through the National Research Foundation of Korea (NRF) funded by the Ministry of Education, Science and Technology(2011-0013021). The material in this paper has been presented in part at the IEEE 79th Vehicular Technology Conference, May 2014, Seoul, Korea.} %
\thanks{Khanh Pham is with the Department of Electrical and Information Engineering, Seoul National University of Science and Technology, 232 Gongneung-ro, Nowon-gu, Seoul, 139-743, Korea. (e-mail: kpham@seoultech.ac.kr)}
\thanks{Kyungchun Lee (corresponding) is with the Department of Electrical and Information Engineering and the Convergence Institute of Biomedical Engineering and Biomaterials, Seoul National University of Science and Technology. (email: kclee@seoultech.ac.kr)}}

\maketitle

\begin{abstract}

This paper proposes a linear interference alignment (IA) scheme which can be used for uplink channels in a general multicell multiuser MIMO cellular network. The proposed scheme aims to align interference caused by signals from a set of transmitters into a subspace which is established by the signals from only a subset of those transmitters, thereby effectively reducing the number of interfering transmitters. The total degrees of freedom (DoF) achievable by the proposed scheme is given in closed-form expression, and a numerical analysis shows that the proposed scheme can achieve the optimal DoF in certain scenarios and provides a higher total DoF than other related schemes in most cases.

\end{abstract}

EDICS: SPC-INTF

\section{Introduction}
 \label{introduction}

Interference management has been an important research subject in the field of mobile communication, where the spectral efficiency is limited mainly by interference \cite{multicell_mimo}. The recent work by \cite{IA_K_IC} has attracted a great deal of attention by demonstrating that a $K$-user interference channel can achieve a total of $K/2$ degrees of freedom (DoF) and thus that each user can obtain half of the interference-free DoF regardless of the number of users, which suggests that interference channels are not necessarily interference-limited. This remarkable result was obtained using the \textit{interference alignment} (IA) technique originated by \cite{Communication_X}.

\subsection{An Overview of IA}

The IA technique is an altruistic transmission method in which signals are transmitted in carefully chosen directions in order to align interference signals, i.e., in order to make interference signals overlapped, as much as possible while preserving the decodability of the desired signals. In the literature, IA has been applied to various scenarios, including the $K$-user interference channel \cite{IA_K_IC}, the $K$-user multiple-input multiple-output (MIMO) interference channel \cite{IA_K_MIMO_IC}, the MIMO X channel \cite{DoF_region_X}, and the wireless X network \cite{IA_DoF_X}. Since it first appeared, IA has not only been thoroughly investigated but also been significantly enhanced; for example, the circumstances under which IA can possibly be applied to MIMO interference channels has been studied \cite{Feasibility}, and it has been shown that a higher total DoF can be achieved if IA is used with symbol extension \cite{DoF_region_X} and that less channel information is required if IA is conducted by iterative algorithms \cite{Feasibility_distributed,approaching_capacity}.

The IA techniques can be classified into two categories: signal-level alignment and signal-space alignment \cite{Feasibility}. The signal-level IA schemes, e.g., the schemes given in \cite{real_IA,IA_on_the_deterministic_channel}, align interferences on the basis of the levels of those interferences; levels can be generally defined as scalar numbers that are independent of each other, e.g., levels were defined in \cite{real_IA} as power levels and in \cite{IA_on_the_deterministic_channel} as the rationally independent products of the beamforming gain and the channel gain. The main benefit of signal-level IA schemes is that they provide a powerful tool to characterize the DoF of a given network; however, they are not suitable for practical systems due to their requirement of a considerable amount of transmit power as they require a very large number of transmit directions.

On the other hand, the signal-space IA schemes not only can be used to characterize the DoF of a given network but can also be adapted to practical systems. The signal-space IA schemes align interferences on the basis of their received directions, which are vectors in the Euclidean space. These schemes can be classified into two categories: asymptotic and linear schemes. The asymptotic schemes, e.g., the schemes given in \cite{IA_K_IC,IA_DoF_X}, require an infinite number of time/frequency/spatial extensions and are mainly used to characterize the DoF of a network, whereas the linear schemes require a finite number of extensions and hence can be used for practical systems.

The signal-space linear IA schemes can be further divided into two types: iterative and non-iterative schemes. The iterative schemes, e.g., the schemes given in \cite{Feasibility_distributed,approaching_capacity}, has been considered to be more practical than the non-iterative schemes as they only require local channel knowledge. However, the number of DoF achievable by the iterative schemes cannot be tractable, whereas that of the non-iterative schemes can. This intractability problem leads to another problem of deciding the optimal number of transmit signals when using the iterative schemes. The iterative schemes also suffer from the problem of the convergence of the IA solutions. Specifically, if this convergence does not exist, only suboptimal IA solutions can be found, and even if the convergence does exist, it is usually slow and thus requires a large number of iterations to approach the IA solution, thereby causing high computational complexity. These problems can be avoided by using the non-iterative schemes, and the only price that has to be paid is the availability of the global channel knowledge.

\subsection{Related Work}

The application of signal-space linear IA for cellular networks were considered in \cite{IA_cellular_network, Two_cell_IFBC, IA_techniques_mimo_ifbd, dof_mimo_ifbc, genie_chain, spatial_dofs_mimo_mac, dof_decomposition, Lee, Hwang, mu_two_way_relay, on_achievability_3_cell, on_feasibility_l_cell}, which used non-iterative schemes, and in \cite{Downlink, Feasibility_algorithm, Iterative_IA_cellular, ia_in_mimo_cell}, which used iterative schemes. Furthermore, the feasibility of non-iterative schemes was studied in \cite{on_feasibility_l_cell,Feasibility_algorithm,ia_in_mimo_cell,on_feasibility_mimo_ifbc}.

The IA scheme given in \cite{IA_cellular_network}, which is called subspace IA, is the first IA scheme developed for cellular networks. The scheme exploits channel decomposibility to align interference into a subspace smaller than the desired signal space and can achieve high total DoF when the number of users is large and the number of cells is small. However, only uplink channels in single-antenna cellular networks were considered in \cite{IA_cellular_network}, and the scheme can only be applied when the channel has the special property of decomposibility, which, for example, is not satisfied by channels with independent identically distributed coefficients \cite{ia_in_mimo_cell}.

The signal-space linear non-iterative IA schemes for downlink channels were considered in \cite{Two_cell_IFBC, IA_techniques_mimo_ifbd, dof_mimo_ifbc, genie_chain}. The scheme in \cite{Two_cell_IFBC} was specifically developed for two-cell two-user MIMO downlink channels, and it used IA to make inter-cell interference (ICI) caused by one base station (BS) at different interfered users to be the same; thus, the overall ICI caused by that BS is effectively reduced. This alignment idea was then used in \cite{IA_techniques_mimo_ifbd, dof_mimo_ifbc} to develop IA schemes for downlink channels in multicell multiuser MIMO networks, whereas the scheme in \cite{genie_chain} used the idea of subspace alignment chain which was introduced in \cite{subspace_alignment_chain}.

The signal-space linear non-iterative IA schemes for uplink channels were considered in \cite{spatial_dofs_mimo_mac, dof_decomposition, Lee, Hwang, mu_two_way_relay, on_achievability_3_cell, on_feasibility_l_cell}. The scheme in \cite{spatial_dofs_mimo_mac} was designed only for two-cell uplink channels in which the transmit and receive antennas satisfy some special constraints, and it was shown that in those special cases the scheme achieved the optimal DoF. For two-cell uplink channels with arbitrary numbers of transmit and receive antennas, even though the structured scheme in \cite{dof_decomposition} can achieve the optimal DoF, it can only be applied when the number of users in each cell is two or three, whereas the scheme in \cite{Lee} can be used for an arbitrary number of users in each cell. The idea used by \cite{Lee} is essentially similar to that used by \cite{Two_cell_IFBC}; that is, IA was used in \cite{Lee} to make ICI caused by users in one cell to be the same at the interfered BS.

For general multicell multiuser uplink channels, \cite{dof_decomposition} proposed a new scheme called unstructured IA scheme which appeared to provide a high DoF performance, but for a given uplink channel the feasibility of the scheme cannot be determined until the scheme is actually applied and checked by a numerical test. On the other hand, the IA scheme in \cite{Hwang} aimed to align ICI into a one-dimensional subspace, and because this IA constraint was too stringent to be satisfied, only a fraction of the ICI could be aligned. This problem was then resolved in \cite{mu_two_way_relay} by consolidating ICI into a multi-dimensional subspace rather than a one-dimensional subspace. By extending the scheme in \cite{on_achievability_3_cell}, which was designed only for three-cell multiuser uplink channels, the scheme in \cite{on_feasibility_l_cell} can be applied to a general uplink channel; however, its DoF performance decreases when the size of the uplink channel increases as the scheme can only suppress the interference caused by one interfering user regardless of the size of the uplink channel.

\subsection{Contributions}

The new signal-space linear non-iterative IA scheme can be used for general multicell multiuser uplink channels in which not only the numbers of cells, users in each cell, transmit antennas, and receive antennas, but also the total number of links from the users in one cell to their interfered receivers, are arbitrary. To further improve the DoF performance of the proposed scheme when it is applied to a specific uplink channel, a new method that can be used to find the best parameters for the proposed scheme is introduced. The closed-form expression for the DoF achieved by the proposed scheme in an uplink channel is also presented.

The basic idea of the proposed IA scheme is to align the ICI caused by a set of users into a subspace that is as large as a subspace generated by the ICI caused by a subset of those users, thereby reducing the dimension of the ICI subspace by a ratio of (the number of users in the set$/$the number of users in the subset). This reduction ratio is similar to the packing ratio, which was introduced in \cite{dof_decomposition} as the ratio of the dimension of the interference subspace before IA to that after IA.

The key feature of the proposed scheme is that it does not explicitly specify the structure of the signal subspace into which the ICI is aligned but does explicitly specify the size of that signal subspace. As a result, when the proposed scheme is applied, the exact structure of the resultant ICI subspace after IA is unknown, whereas its size is known precisely; that is, the resultant ICI subspace is as large as the subspace created by the ICI caused by users in the subset.

Owing to this key feature, the design of the proposed scheme can focus on the final outcome of the IA rather than on the details of the IA, simplifying the design itself. The design approach used by the proposed scheme is quite similar to those used by the schemes given in \cite{dof_decomposition,on_achievability_3_cell,on_feasibility_l_cell} but differs from those used by the schemes given in \cite{Lee, Hwang, mu_two_way_relay}, which try to align interference into a specific subspace. Because of their strict requirements, the IA schemes in \cite{Lee, Hwang, mu_two_way_relay} either cannot be used for a general uplink channel or leave a large amount of interference unaligned; in contrast, the proposed scheme does not suffer from these problems, so it can provide more DoF than the schemes in \cite{Lee, Hwang, mu_two_way_relay}.

Despite the similarity between the approaches used by the proposed scheme and the schemes in \cite{dof_decomposition,on_achievability_3_cell,on_feasibility_l_cell}, the proposed scheme still provides some advantages over the other schemes in terms of the DoF performance and computational complexity. Specifically, at each BS, the IA schemes in \cite{on_achievability_3_cell,on_feasibility_l_cell} can eliminate only one inter-cell interfering user, whereas the proposed scheme can reduce the number of inter-cell interfering users by a ratio of (the number of users in the set$/$the number of users in the subset); therefore, the proposed scheme can provide a larger space for the desired signals and hence can achieve more DoF. On the other hand, compared to the scheme in \cite{dof_decomposition}, the proposed scheme not only requires a considerably lower complexity (as it performs IA only on the inter-cell interfering users from the same cell, whereas the scheme in \cite{dof_decomposition} performs IA on all the inter-cell interfering streams) but can also provide more DoF in certain scenarios. Furthermore, the proposed scheme can guarantee the decodability of the desired signals with high probability, whereas the scheme in \cite{dof_decomposition} cannot.

This paper is organized as follows. In Section \ref{sec:system_model}, we present the system model of an uplink channel in a multicell multiuser MIMO network and the DoF metric that will be used to evaluate the performance of the proposed scheme. In Section \ref{sec:proposed_scheme}, we develop the proposed IA scheme for a simple scenario, and then in Section \ref{sec:application_proposed_scheme}, we present the application of the proposed scheme in the uplink channel described in Section \ref{sec:system_model}. In Section \ref{sec:numerical_analysis}, we compare the DoF performance of the proposed scheme to the optimal DoF and those of other non-iterative signal-space IA schemes. Finally, in Section \ref{sec:conclusion}, we conclude the paper.

\textit{Notation:} Throughout this paper, the following notation will be used. $\mathcal{C}(\mathbf{A}_{1},\mathbf{A}_{2},\cdots,\mathbf{A}_{n})$ and $\mathcal{N}(\mathbf{A})$ denote the column space of matrix $\big[ \mathbf{A}_{1} \ \mathbf{A}_{2} \ \cdots \ \mathbf{A}_{n} \big]$ and the null space of matrix $\mathbf{A}$, respectively. $\text{dim}\big( \mathcal{C}( \mathbf{A} ) \big)$ and $\text{rank}( \mathbf{A} )$ denote the dimension of $\mathcal{C}( \mathbf{A} )$ and the rank of matrix $\mathbf{A}$, respectively. $\mathbf{I}_{n}$ and $\text{diag}\big( \mathbf{A}_{1},\mathbf{A}_{2},\cdots,\mathbf{A}_{n} \big)$ represent the $n \times n$ identity matrix and the block diagonal matrix whose $i$th block is the matrix $\mathbf{A}_{i}$, respectively. In order to simplify the representation, $\mathbf{0}$ is used to denote the zero matrix of any size. Finally, the symbols \textquotedblleft$\equiv$\textquotedblright, \textquotedblleft$\setminus$\textquotedblright, and \textquotedblleft$T$\textquotedblright \ denote the equivalent, set minus, and transpose operations, respectively.

\section{System Model and DoF} \label{sec:system_model}

In this section, we present the system model of an uplink channel in a multicell multiuser MIMO network and give a brief review on the DoF metric.

\subsection{System Model}

The main system considered in this paper is an uplink channel in a cellular network consisting of $L$ cells, each having one BS serving $K$ users. Each BS has $M_{r}$ receive antennas, while each user has $N_{t}$ transmit antennas and transmits $d$ data streams.

As each BS receives signals from all users in the network, the signal vector received at BS $i$ is given by
\begin{equation} \label{eq:system_model}
 \mathbf{y}_{i} = \underbrace{ \sum_{k=1}^{K} \mathbf{H}_{i}^{ik} \mathbf{V}^{ik} \mathbf{s}^{ik} }_{\text{Desired Signals}} + \underbrace{ \sum_{\substack{ l = 1 \\ l \neq i }}^{L} \sum_{k=1}^{K}  \mathbf{H}_{i}^{lk} \mathbf{V}^{lk} \mathbf{s}^{lk} }_{\text{Inter-cell interference}} + \mathbf{z}_{i},
\end{equation}
where $\mathbf{H}_{i}^{lk} \in \mathbb{C}^{M_r \times N_t}$ is the channel matrix between BS $i$ and user $k$ in cell $l$, $\mathbf{V}^{lk} \in \mathbb{C}^{N_t \times d}$ is the precoder matrix used by user $k$ in cell $l$, $\mathbf{s}^{lk} \in \mathbb{C}^{d \times 1}$ is the data vector that user $k$ in cell $l$ transmits to its serving BS, and $\mathbf{z}_{i} \in \mathbb{C}^{M_{r} \times 1}$ is a zero mean unit variance circularly symmetric additive white Gaussian noise vector at BS $i$. It is assumed that all channel matrices are known and that the channel is Rayleigh fading; hence, the coefficients in the channel matrices are complex independent identically Gaussian distributed random variables with zero mean and unit variance.

Owing to the average power constraint in cellular networks, the precoder and data vector from user $k$ in cell $i$ are normalized such that the average transmit power of the user is limited by $P^{ik}$, i.e., $\mathbf{E} \big[ || \mathbf{V}^{ik}\mathbf{s}^{ik} ||^{2} \big] \leq P^{ik}$, where $\mathbf{E}[.]$ and $||.||$ are respectively the expectation and norm functions, and $P^{ik}$ is the maximum transmit power for user $k$ in cell $i$. This average power constraint can be guaranteed by selecting the precoder and data vector satisfying
\[
\left\lbrace
\begin{array}{l}
\mathbf{E}\big[ \mathbf{s}^{ik} ( \mathbf{s}^{ik} )^{H} \big] = \text{diag}( p_{1}^{ik}, p_{2}^{ik}, \cdots, p_{d}^{ik} ), \\
|| \mathbf{v}_{j}^{ik} ||^2 = 1, \ j = 1,2,\cdots,d,
\end{array}
\right.
\]
where $p_{j}^{ik}$ is the transmit power for the $j$-th signal stream from user $k$ in cell $i$ and satisfies $\sum_{j=1}^{d} p_{j}^{ik} \leq P^{ik}$, and $\mathbf{v}_{j}^{ik}$ is the $j$-th column in $\mathbf{V}^{ik}$.

\subsection{Degree of Freedom}

In high-signal-to-noise-ratio regimes, the DoF of one user can be numerically calculated from the data rate of that user. Specifically, if $d^{ik}$ is the DoF of user $k$ in cell $i$, then the data rate of that user is expressed as
\[
R^{ik} = d^{ik}\log_{2}(SNR^{ik}) + o\big( \log_{2}(SNR^{ik}) \big),
\]
where $SNR^{ik}$ is the signal-to-noise ratio corresponding to user $k$ in cell $i$, and $SNR^{ik} = P^{ik}$ as the noise has unit power. When $P^{ik}$ is large, $d^{ik}\log_{2}(P^{ik})$ approaches $R^{ik}$ as the contribution of $o\big( \log_{2}(P^{ik}) \big)$ becomes negligible \cite{IA_K_IC}. Thus, the DoF of user $k$ in cell $i$ is given by
\[
d^{ik} = \lim_{P^{ik} \rightarrow \infty} \frac{ R^{ik} }{ \log_{2}( P^{ik} ) }.
\]

The DoF of one user can alternatively be interpreted as the number of data streams transmitted by that user and decodable by the desired receiver of that user \cite{IA_DoF_X}. The decodability is achieved when the transmitted streams are received in the directions independent of the directions on which the other signals are received. Therefore, each user in the uplink channel of \eqref{eq:system_model} achieves $d$ DoF if at each BS the desired signals are both independent of each other and independent of the ICI signals.

\section{The proposed IA scheme} \label{sec:proposed_scheme}

\begin{figure}[h]
	\centering
	\includegraphics[trim= 50 290 10 60, clip=true, scale=0.36]{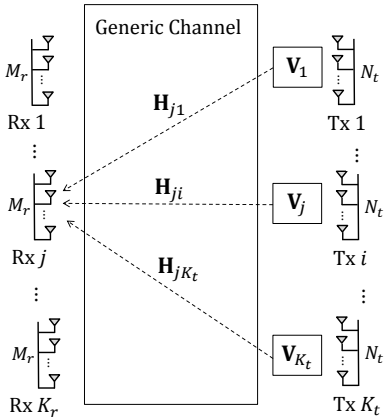}
	\caption{Interference signals received at receiver $j$ in the specific scenario used in Section \ref{sec:proposed_scheme} to develop the proposed scheme.}
	\label{fig:system_model_2}
\end{figure}

In this section, the proposed IA scheme is presented. The proposed scheme is developed in this section for a different scenario from the uplink channel given in \eqref{eq:system_model}. Specifically, the scenario considered in this section consists of $K_{t}$ $N_{t}$-antenna transmitters that cause interference to the same $K_{r}$ $M_{r}$-antenna receivers. Furthermore, with an abuse of the notation, this section uses $\mathbf{H}_{ji} \in \mathbb{C}^{M_{r} \times N_{t}}$ and $\mathbf{V}_{i} \in \mathbb{C}^{N_{t} \times d}$ to denote the channel matrix from transmitter $i$ to receiver $j$ and the precoder of transmitter $i$, respectively. The scenario is depicted in Fig. \ref{fig:system_model_2}.

The proposed IA scheme aims to jointly design the precoders of the $K_{t}$ users so that the following two goals are satisfied: (1) the interference caused by the $K_{t}$ users at the same receiver is aligned into a subspace spanned by the interference caused by $\kappa_{t}$ users among those $K_{t}$ users, and (2) the precoders are of full column rank. By doing so, the number of interference signals is reduced by a ratio of $K_{t}/\kappa_{t}$, while the signals of those $K_{t}$ users are guaranteed to be decodable by their desired receiver.

The basic idea of the proposed IA scheme is to design the precoders on the basis of the following lemma:

\newcounter{tempCounter}
\begin{figure*}[t]
	\setcounter{tempCounter}{\value{equation}}
	\setcounter{equation}{4}
	\begin{equation} \label{eq:design_precoders_for_users_in_one_cell_extended}
		\begin{array}{cccc}
			K_{r}( K_{t} - \kappa_{t} )M_{r} \underbrace{ \left\{ \left[
				\begin{array}{cccc}
					\gamma_{11}^{1} \mathbf{H}_{11} & \gamma_{12}^{1} \mathbf{H}_{12} & \cdots & \gamma_{1K_{t}} \mathbf{H}_{1K_{t}} \\
					\gamma_{21}^{1} \mathbf{H}_{11} & \gamma_{22}^{1} \mathbf{H}_{12} & \cdots & \gamma_{2K_{t}}^{1} \mathbf{H}_{1K_{t}} \\
					\vdots & \vdots & \ddots & \vdots \\
					\gamma_{(K_{t} - \kappa_{t})1}^{1} \mathbf{H}_{11} & \gamma_{(K_{t} - \kappa_{t})2}^{1} \mathbf{H}_{12} & \cdots & \gamma_{(K_{t} - \kappa_{t})K_{t}}^{1} \mathbf{H}_{1K_{t}} \\
					\gamma_{11}^{2} \mathbf{H}_{21} & \gamma_{12}^{2} \mathbf{H}_{22} & \cdots & \gamma_{1K_{t}}^{2} \mathbf{H}_{2K_{t}} \\
					\gamma_{21}^{2} \mathbf{H}_{21} & \gamma_{22}^{2} \mathbf{H}_{22} & \cdots & \gamma_{2K_{t}}^{2} \mathbf{H}_{2K_{t}} \\
					\vdots & \vdots & \ddots & \vdots \\
					\gamma_{(K_{t} - \kappa_{t})1}^{2} \mathbf{H}_{21} & \gamma_{(K_{t} - \kappa_{t})2}^{2} \mathbf{H}_{22} & \cdots & \gamma_{(K_{t} - \kappa_{t})K_{t}}^{2} \mathbf{H}_{2K_{t}} \\
					\vdots & \vdots & \ddots & \vdots \\
					\gamma_{11}^{K_{r}} \mathbf{H}_{K_{r}1} & \gamma_{12}^{K_{r}} \mathbf{H}_{K_{r}2} & \cdots & \gamma_{1K_{t}}^{K_{r}} \mathbf{H}_{K_{r}K_{t}} \\
					\gamma_{21}^{K_{r}} \mathbf{H}_{K_{r}1} & \gamma_{22}^{K_{r}} \mathbf{H}_{K_{r}2} & \cdots & \gamma_{2K_{t}}^{K_{r}} \mathbf{H}_{K_{r}K_{t}} \\
					\vdots & \vdots & \ddots & \vdots \\
					\gamma_{(K_{t} - \kappa_{t})1}^{K_{r}} \mathbf{H}_{K_{r}1} & \gamma_{(K_{t} - \kappa_{t})2}^{K_{r}} \mathbf{H}_{K_{r}2} & \cdots & \gamma_{(K_{t} - \kappa_{t})K_{t}}^{K_{r}} \mathbf{H}_{K_{r}K_{t}}
				\end{array}
			\right] \right. }_{K_{t}N_{t}}
			&
			\left[
				\begin{array}{c}
					\mathbf{V}_{1} \\
					\mathbf{V}_{2} \\
					\vdots \\
					\mathbf{V}_{K_{t}}
				\end{array}
			\right]
			&
			=
			&
			\mathbf{0}.
		\end{array}
	\end{equation}
	\setcounter{equation}{\value{tempCounter}}
	\hrulefill
	\vspace{4pt}
\end{figure*}

\begin{lemma} \label{lemma_randomized_sum}
Given $\big\{ \mathbf{A}_{k}: \mathbf{A}_{k} \in \mathbb{C}^{m_{1} \times m_{2}}$, $\mbox{rank}(\mathbf{A}_{k}) \leq r$, $k = 1,2,\cdots,n_{1} \big\}$; if we have
\begin{multline} \label{eq:lemma3}
\gamma_{i1}\mathbf{A}_1 + \gamma_{i2}\mathbf{A}_2 + \cdots + \gamma_{in_{1}}\mathbf{A}_{n_{1}} = \mathbf{0},\\
\forall i = 1,2,\cdots,n_{1}-n_{2},
\end{multline}
where $\gamma_{ik},\ k=1,2,\cdots,n_{1}, \ i = 1,2,\cdots,n_{1}-n_{2}$ are independent identically distributed (i.i.d.) scalars and $0 \leq n_{2} \leq n_{1}$, then
\[
\text{dim} \Big( \mathcal{C} \big( \mathbf{A}_1, \ \mathbf{A}_2, \ \cdots, \ \mathbf{A}_{n_{1}} \big) \Big) \leq n_{2} r.
\]
The equality occurs when $n_{2}$ subspaces among $n_{1}$ subspaces $\mathcal{C} (\mathbf{A}_{k}), \ k=1,2,\cdots,n_{1}$ are of dimension $r$ and disjoint, and $m_{1} \geq n_{2} r$.
\end{lemma}
\begin{IEEEproof}
See Appendix \ref{proof_lemma_randomized_sum}.
\end{IEEEproof}
The lemma basically states that if \eqref{eq:lemma3} is satisfied, then the column vectors of $n_{1}$ matrices $\mathbf{A}_{k}, \ k=1,2,\cdots,n_{1}$ are aligned into a subspace spanned by the column vectors of $n_{2}$ matrices among those $n_{1}$ matrices.

Following \textit{Lemma \ref{lemma_randomized_sum}}, the first goal can be achieved by designing the precoders for the $K_{t}$ users such that
\begin{multline} \label{eq:design_precoders_for_users_in_one_cell}
\gamma_{i1}^{j} \mathbf{H}_{j1} \mathbf{V}_{1} + \gamma_{i2}^{j} \mathbf{H}_{j2} \mathbf{V}_{2} + \cdots + \gamma_{iK_{t}}^{j} \mathbf{H}_{jK_{t}} \mathbf{V}_{K_{t}}
= \mathbf{0},\\
\forall i=1,2,\cdots,K_{t}-\kappa_{t},\ \forall j = 1,2,\cdots,K_{r},
\end{multline}
where $\gamma_{ik}^{j}, \ \forall i,j,k$ are i.i.d. scalars.

As the precoder selection is constrained by the first goal, which requires that the precoders satisfy \eqref{eq:design_precoders_for_users_in_one_cell}, the second goal can be achieved only if the precoders found by solving  \eqref{eq:design_precoders_for_users_in_one_cell} are also of full column rank. In the following, we will show that these precoders are in fact of full column rank when
\begin{equation} \label{eq:constraint_kappa}
(K_{t} - 1) N_{t} \leq K_{r} (K_{t} - \kappa_{t}) M_{r} < K_{t} N_{t}.
\end{equation}

Let us first rewrite \eqref{eq:design_precoders_for_users_in_one_cell} as \eqref{eq:design_precoders_for_users_in_one_cell_extended}, which is given at the top of the next page. As $\big[ \mathbf{V}_{1}^{T},\mathbf{V}_{2}^{T},\cdots,\mathbf{V}_{K_{t}}^{T} \big]^{T}$ is the solution to \eqref{eq:design_precoders_for_users_in_one_cell_extended}, the necessary condition for $\mathbf{V}_{i}, \ i=1,2,\cdots,K_{t}$ to be of full column rank is that $\text{dim}\big( \mathcal{N}( \mathbf{C} ) \big) \geq d$, where $\mathbf{C}$ is the coefficient matrix in \eqref{eq:design_precoders_for_users_in_one_cell_extended}. According to \textit{Lemma \ref{lemma:rank_coef_matrix}}, which is given in Appendix \ref{app:rank_coef_matrix}, if $N_{t} \leq K_{r}M_{r}$, $\mathbf{C}$ is a full rank matrix, which leads to $\text{dim}\big( \mathcal{N}( \mathbf{C} ) \big) = K_{t}N_{t} - K_{r}(K_{t} - \kappa_{t})M_{r}$. Consequently, given that $N_{t} \leq K_{r}M_{r}$, the necessary condition of $\text{dim}\big( \mathcal{N}( \mathbf{C} ) \big) \geq d$ can always be satisfied by choosing $d$ such that
\addtocounter{equation}{1}
\begin{equation} \label{eq:necessary_condition}
K_{t}N_{t} - K_{r}(K_{t} - \kappa_{t}) M_{r} \geq d.
\end{equation}

Given the constraint \eqref{eq:constraint_kappa} and the necessary condition \eqref{eq:necessary_condition}, the proof that the precoders of the $K_{t}$ users are of full column rank is presented in detail in Appendix \ref{app:proof_full_rank_precoders}. Here we give a simple example to illustrate the basic idea of the proof; in the example, we assume that $K_{r} = 1$, $K_{t} = 3$, $\kappa_{t} = 1$, $M_{r} = N_{t}$, and $d = 2$. The proof is then given as follows.

We first assume the contradiction that $\mathbf{V}_{1}$ is not a full column rank matrix; then there exist two scalars $\{ \lambda_{i},\ i=1,2: \exists \lambda_{i} \neq 0 \}$ satisfying
\begin{equation} \label{eq:simple_example_contradiction_first_precoder}
\lambda_{1}\mathbf{v}_{1,1} + \lambda_{2}\mathbf{v}_{1,2} = \mathbf{0},
\end{equation}
where $\mathbf{v}_{i,j}$ is the $j$-th column in $\mathbf{V}_{i}$.

The IA system of \eqref{eq:design_precoders_for_users_in_one_cell_extended} for this example is given by
\begin{equation} \label{eq:simple_example_proof_full_rank_precoders}
\setlength\arraycolsep{3pt}
\begin{array}{cccc}
	\begin{bmatrix}
	\gamma_{11}^{1} \mathbf{H}_{11} & \gamma_{12}^{1} \mathbf{H}_{12} & \gamma_{13}^{1} \mathbf{H}_{13} \\
	\gamma_{21}^{1} \mathbf{H}_{11} & \gamma_{22}^{1} \mathbf{H}_{12} & \gamma_{23}^{1} \mathbf{H}_{13} \\	
	\end{bmatrix}
	&
	\begin{bmatrix}
	\mathbf{v}_{1,1} & \mathbf{v}_{1,2} \\
	\mathbf{v}_{2,1} & \mathbf{v}_{2,2} \\
	\mathbf{v}_{3,1} & \mathbf{v}_{3,2}
	\end{bmatrix}
	&
	=
	&
	\mathbf{0},
\end{array}
\end{equation}
which then leads to
\[
\setlength\arraycolsep{3pt}
\begin{array}{cccc}
	\begin{bmatrix}
	\gamma_{11}^{1} \mathbf{H}_{11} & \gamma_{12}^{1} \mathbf{H}_{12} & \gamma_{13}^{1} \mathbf{H}_{13} \\
	\gamma_{21}^{1} \mathbf{H}_{11} & \gamma_{22}^{1} \mathbf{H}_{12} & \gamma_{23}^{1} \mathbf{H}_{13} \\	
	\end{bmatrix}
	&
	\begin{bmatrix}
	\lambda_{1} \mathbf{v}_{1,1} + \lambda_{2} \mathbf{v}_{1,2} \\
	\lambda_{1} \mathbf{v}_{2,1} + \lambda_{2} \mathbf{v}_{2,2} \\
	\lambda_{1} \mathbf{v}_{3,1} + \lambda_{2} \mathbf{v}_{3,2}
	\end{bmatrix}
	&
	=
	&
	\mathbf{0}.
\end{array}
\]
According to \eqref{eq:simple_example_contradiction_first_precoder}, we have
\[
\setlength\arraycolsep{3pt}
\begin{array}{cccc}
	\begin{bmatrix}
	\gamma_{11}^{1} \mathbf{H}_{11} & \gamma_{12}^{1} \mathbf{H}_{12} & \gamma_{13}^{1} \mathbf{H}_{13} \\
	\gamma_{21}^{1} \mathbf{H}_{11} & \gamma_{22}^{1} \mathbf{H}_{12} & \gamma_{23}^{1} \mathbf{H}_{13} \\	
	\end{bmatrix}
	&
	\begin{bmatrix}
	\mathbf{0} \\
	\lambda_{1} \mathbf{v}_{2,1} + \lambda_{2} \mathbf{v}_{2,2} \\
	\lambda_{1} \mathbf{v}_{3,1} + \lambda_{2} \mathbf{v}_{3,2}
	\end{bmatrix}
	&
	=
	&
	\mathbf{0},
\end{array}
\]
and hence
\[
\setlength\arraycolsep{3pt}
\begin{array}{cccc}
	\mathbf{C}_{2}
	&
	\begin{bmatrix}
	\lambda_{1} \mathbf{v}_{2,1} + \lambda_{2} \mathbf{v}_{2,2} \\
	\lambda_{1} \mathbf{v}_{3,1} + \lambda_{2} \mathbf{v}_{3,2}
	\end{bmatrix}
	&
	=
	&
	\mathbf{0},
\end{array}
\]
where
\[
\mathbf{C}_{2} = \begin{bmatrix}
	\gamma_{12}^{1} \mathbf{H}_{12} & \gamma_{13}^{1} \mathbf{H}_{13} \\
	\gamma_{22}^{1} \mathbf{H}_{12} & \gamma_{23}^{1} \mathbf{H}_{13} \\	
	\end{bmatrix}.
\]
Therefore, either $\big[ (\lambda_{1} \mathbf{v}_{2,1} + \lambda_{2} \mathbf{v}_{2,2})^{T}, (\lambda_{1} \mathbf{v}_{3,1} + \lambda_{2} \mathbf{v}_{3,2})^{T} \big]^{T}$ is a non-zero vector and belongs to $\mathcal{N}(\mathbf{C_{2}})$, or $\big[ (\lambda_{1} \mathbf{v}_{2,1} + \lambda_{2} \mathbf{v}_{2,2})^{T}, (\lambda_{1} \mathbf{v}_{3,1} + \lambda_{2} \mathbf{v}_{3,2})^{T} \big]^{T} = \mathbf{0}$. The first possibility cannot be the case, as $\mathbf{C}_{2}$ is a full column rank matrix according to \textit{Lemma \ref{lemma:rank_coef_matrix}} and the constraint \eqref{eq:constraint_kappa}. On the other hand, according to the necessary condition \eqref{eq:necessary_condition}, the solution to \eqref{eq:simple_example_proof_full_rank_precoders} can always be chosen to be of full column rank; thus, the second possibility cannot be the case, as it will lead to
\[
\setlength\arraycolsep{1.2pt}
\begin{array}{ccccc}	
	\begin{bmatrix}
	\lambda_{1} \mathbf{v}_{1,1} + \lambda_{2} \mathbf{v}_{1,2} \\
	\lambda_{1} \mathbf{v}_{2,1} + \lambda_{2} \mathbf{v}_{2,2} \\
	\lambda_{1} \mathbf{v}_{3,1} + \lambda_{2} \mathbf{v}_{3,2}
	\end{bmatrix}
	&
	=
	&
	\lambda_{1}
	\begin{bmatrix}	
	\mathbf{v}_{1,1} \\
	\mathbf{v}_{2,1}\\
	\mathbf{v}_{3,1}
	\end{bmatrix} +
	\lambda_{2}
	\begin{bmatrix}	
	\mathbf{v}_{1,2} \\
	\mathbf{v}_{2,2} \\
	\mathbf{v}_{3,2}
	\end{bmatrix}
	&
	=
	&
	\mathbf{0},
\end{array}
\]
which implies that the solution to \eqref{eq:simple_example_proof_full_rank_precoders} is not of full column rank. Because the consequences of the assumed contradiction \eqref{eq:simple_example_contradiction_first_precoder} are impossible, we conclude that \eqref{eq:simple_example_contradiction_first_precoder} cannot occur, and hence $\mathbf{V}_{1}$ is of full column rank. Similarly, we can show that $\mathbf{V}_{k}, \ k = 2,3,\cdots,K_{t}$ are also of full column rank.

To summarize, the outcomes of the proposed IA scheme are given in the following proposition:
\begin{proposition} \label{proposition1}
Given a scenario where $K_{t}$ $N_{t}$-antenna transmitters cause interference to the same $K_{r}$ $M_{r}$-antenna receivers and $N_{t} \leq K_{r} M_{r}$, the proposed IA scheme can design the precoders for $K_{t}$ users such that
\[
\left\lbrace
\begin{array}{l}
	\mathbf{V}_{k} \in \mathbb{C}^{N_{t} \times d} \text{ is of full column rank}, \ \forall k = 1,2,\cdots,K_{t}, \\
	\text{dim}\Big( \mathcal{C} \big( \mathbf{H}_{j1} \mathbf{V}_{1}, \mathbf{H}_{j2} \mathbf{V}_{2}, \cdots, \mathbf{H}_{jK_{t}} \mathbf{V}_{K_{t}} \big) \Big) \leq \kappa_{t} d, \\
	\qquad\qquad\qquad\qquad\qquad\qquad\qquad\qquad \forall j = 1,2,\cdots,K_{r},
\end{array}
\right.
\]
where $\kappa_{t}$ and $d$ satisfy \eqref{eq:constraint_kappa} and \eqref{eq:necessary_condition}, respectively.
\end{proposition}

\textit{Remark:} When $N_{t} > K_{r} M_{r}$, the interference caused by the $K_{t}$ users can be completely eliminated by selecting $\mathbf{V}_{k}, \ k = 1,2,\cdots,K_{t}$ from the nullspace of the matrix
\[
 \bar{\mathbf{H}}_{k} = \big[ \mathbf{H}_{1k}^{T}, \mathbf{H}_{2k}^{T}, \cdots,\mathbf{H}_{K_{r}k}^{T} \big]^{T}.
\]
As $\bar{\mathbf{H}}_{k} \in \mathbb{C}^{K_{r}M_{r} \times N_{t}}$ consists of i.i.d. entries and $N_{t} > K_{r}M_{r}$, a non-trivial nullspace of $\bar{\mathbf{H}}_{k}$ always exists, and thus a full column rank $\mathbf{V}_{k}$ can be found.

From the development of the proposed scheme, we can see that it can be applied to a variety of networks where the channel coefficients are generic and there are $K_{t}$ transmitters causing interference to the same $K_{r}$ receivers. Examples of these networks include uplink channels in multicell multiuser MIMO networks, partially connected uplink channels \cite{partially_connected}, and Z-interference uplink channels \cite{Lee}. In the next section, we will present in detail the application of the proposed scheme to an uplink channel in a multicell multiuser cellular network.

\section{The Application of the Proposed IA Scheme in Multicell Multiuser Uplink MIMO Channels} \label{sec:application_proposed_scheme}

In this section, we show how the proposed IA scheme is applied to the uplink channel given in \eqref{eq:system_model}. In the first subsection, we develop the method of determining $K_{t}$ and $K_{r}$ so that the proposed scheme can achieve the best DoF performance, and we also present the closed-form expression for the achievable DoF of the proposed scheme. In the second subsection, we describe in detail the application of the proposed scheme in the uplink channel. Finally, in the third subsection, we derive the upper bound of the achievable DoF of the proposed scheme in the uplink channel.

\subsection{Method of determining the best values of $K_{t}$ and $K_{r}$}

As the goal of the method is to find the values of $K_{t}$ and $K_{r}$ that produce the best DoF performance for the proposed scheme, we need to solve two problems when developing the method. First, we need to determine all the possible values of $K_{t}$ and $K_{r}$. Second, we need to establish the closed-form expression for the achievable DoF, which enables us to determine which values of $K_{t}$ and $K_{r}$ yield the best DoF performance.

Regarding the first problem, we notice that in the uplink channel given in \eqref{eq:system_model}, all the users in the same cell cause interference to the same receivers. Therefore, as there is a total of $K$ users in each cell and a total of $(L-1)$ BSs to which the $K$ users cause interference, the values of $K_{t}$ and $K_{r}$ fall in the intervals of $[ 1,K ]$ and $[ 1,L-1 ]$, respectively; i.e.,
\[
\left\lbrace
\begin{array}{l}	
	1 \leq K_{t} \leq K, \\
	1 \leq K_{r} \leq L-1.
\end{array}
\right.
\]

To handle the second problem, we need to find the number of transmitted signals that can be decoded by their desired receivers, i.e., the number of decodable transmitted signals. This number depends on the number of transmit directions that can be found by the proposed scheme, the amount of interference caused by these transmit directions, and the size of the signal space at each receiver.

As the proposed scheme designs the transmit directions by solving the IA systems having a form similar to \eqref{eq:design_precoders_for_users_in_one_cell_extended}, the number of transmit directions that it can find depends on both the number of IA systems and the number of transmit directions selected from one IA system. Because for each set of $K_{t}$ users we can establish one IA system, the maximum number of IA systems that contain one common user is $\binom{K-1}{K_{t}-1}$. For example, let us assume that $K = 4$ and $K_{t} = 3$; then the transmit directions of user 1 can be found by solving $\binom{3}{2} = 3$ IA systems formed by the following sets of users: $\{ 1,2,3 \},\ \{ 1,2,4 \},$ and $\{ 1,3,4 \}$. On the other hand, the number of transmit directions of one user found by solving one IA system needs to satisfy the necessary condition of \eqref{eq:necessary_condition} in order for those transmit directions to be linearly independent of each other. Therefore, the maximum number of transmit directions of one user that can be found by solving all possible IA systems is given by
\[
\binom{K-1}{K_{t}-1} \big( K_{t} N_{t} - K_{r} (K_{t} - \kappa_{t}) M_{r} \big),
\]
which leads to
\begin{equation} \label{eq:constraint_n_transmit_directions}
d \leq \binom{K-1}{K_{t}-1} \big( K_{t} N_{t} - K_{r} (K_{t} - \kappa_{t}) M_{r} \big).
\end{equation}

Note that even though the transmit directions of one user might be found from different IA systems, they are still independent of each other. This is because the transmit directions found from one system are independent of each other owing to the nature of the proposed scheme, and the transmit directions found from different systems are also independent of each other, as the channel is generic and all the IA systems are different \cite{dof_decomposition}.

The amount of interference caused by the transmitted signals depends on both the number of transmit directions and the efficiency of the proposed IA scheme. When $K_{r} < L$, the IAs at each cell, i.e., the IAs for $K$ users in each cell, do not consider all the interfered BSs; thus, two types of ICI signals exist at each BS: one consists of the ICI signals remaining after IA, and one consists of those that are not involved in any IA. Because the IAs at each cell involve $K_{r}$ interfered BSs and produce $K d$ transmit directions, the total number of ICI signals involved in the IAs at each cell is given by $K_{r} K d$. Owing to the efficiency of the proposed scheme, the number of these ICI signals is effectively reduced by a ratio of $K_{t}/\kappa_{t}$. Therefore, the total number of ICI signals remaining after IA at all the BSs is
\begin{equation} \label{eq:n_ici_signals_after_IA}
L K_{r} K d  (\kappa_{t}/K_{t}).
\end{equation}
On the other hand, as the IAs at each cell do not consider $( L - 1 - K_{r} )$ interfered BSs, the total number of ICI signals that are not involved in any IA is
\begin{equation} \label{eq:n_ici_signals_not_involved_IA}
L ( L - 1 - K_{r} ) K d.
\end{equation}

Because the network is symmetric, it is always possible to select $K_{r}$ interfered BSs for the IAs at each cell so that the numbers of ICI signals at different BSs are equivalent to each other. Thus, from \eqref{eq:n_ici_signals_after_IA} and \eqref{eq:n_ici_signals_not_involved_IA}, the total number of ICI signals at each BS is given by
\[
n_{\text{ICI}} = K_{r} K d (\kappa_{t}/K_{t}) + ( L - 1 - K_{r} ) K d.
\]

The transmitted signals are decodable when the transmit directions are linearly independent of each other and the subspaces created by the desired signals and the ICI signals are separated at each BS. In the proposed scheme, the first constraint is satisfied. On the other hand, as the uplink channel given in \eqref{eq:system_model} is generic and the direct channel matrices are not involved in any IA system, the separability is almost surely satisfied as long as the signal space at each BS is large enough to accommodate both the desired signal subspace and the ICI subspace \cite{dof_decomposition}. Therefore, when the proposed scheme is used, the transmitted signals are decodable if
\begin{equation} \label{eq:enhancements_decodability}
M_{r} \geq K d + n_{\text{ICI}}.
\end{equation}

Consequently, when each user transmits $d$ signals, the decodability of all the transmitted signals is guaranteed by the constraints \eqref{eq:constraint_kappa}, \eqref{eq:constraint_n_transmit_directions}, and \eqref{eq:enhancements_decodability}. Thus, the closed-form expression for the DoF achieved by each user is given by
\begin{multline} \label{eq:achievable_dof}
d = \text{min} \Big( \frac{ M_{r} }{ K + K_{r} K (\kappa_{t}/K_{t}) + ( L - 1 - K_{r} ) K }, \\
\binom{K-1}{K_{t}-1} \big( K_{t} N_{t} - K_{r} (K_{t} - \kappa_{t}) M_{r} \big) \Big),
\end{multline}
in which $\kappa_{t}$ satisfies \eqref{eq:constraint_kappa}.

When $M_{r} > K d + n_{\text{ICI}}$, the total achievable DoF can be further enhanced because the signal space at each BS is not completely filled. The idea is to reuse the proposed scheme to design the new set of precoders to fill that remaining space. Two issues need to be addressed when the proposed scheme is reused. First, the dimension of the available space at each BS is not the number of receive antennas $M_{r}$ but the dimension of the remaining space. Second, the already-designed and to-be-designed precoders belonging to the same user need to be independent of each other. This issue can be resolved by designing the new precoders with respect to the effective channel matrices formed by right-multiplying the channel matrices by the matrices orthogonal to the already-designed precoders \cite{dof_decomposition}. This guarantees that the new precoders belong to the subspace orthogonal to the old precoders; however, the drawback is that the number of transmit antennas is reduced to the dimension of the subspace orthogonal to the already-designed precoders.

To summarize, the basic steps of the method used to determine the best values of $K_{t}$ and $K_{r}$ are presented in Table \ref{tab:dof_enhancement}. In the algorithm, the parameters $\eta_{r}$ and $\eta_{t}$ are the numbers of available receive directions at each BS and available transmit directions at each user, respectively. Hence, in the first run of the algorithm, $\eta_{r}$ and $\eta_{t}$ are set to $M_{r}$ and $N_{t}$, respectively; subsequently, they are set to the dimensions of the remaining space at each BS and the subspace orthogonal to the already-designed precoders, respectively.

\begin{table}[h]
	\centering
	\caption{Method of determining the best values of $K_{t}$ and $K_{r}$}
	\label{tab:dof_enhancement}
	\hrule
	\vspace{\baselineskip}
	\begin{algorithmic}[1]
		\REQUIRE $L$, $K$, $M_{r}$, $\eta_{r}$, $\eta_{t}$
		\STATE $N_{t} := \eta_{t}$
		\FOR{$K_{r} := 1$ to $(L-1)$}
			\FOR{$K_{t} := 1$ to $K$}
				\IF{there exists $\kappa_{t}$ satisfying \eqref{eq:constraint_kappa}}
					\STATE Calculate $d$ using \eqref{eq:achievable_dof}, but replace $M_{r}$ in the first term of the min function with $\eta_{r}$.
					\IF{$\eta_{r} > Kn_{t} + n_{\text{ICI}}$}
						\STATE $\eta_{r} := \eta_{r} - Kn_{t} - n_{\text{ICI}}$
						\STATE $\eta_{t} := \eta_{t} - d$
						\STATE Run this algorithm with the new values of $\eta_{r}$ and $\eta_{t}$; then save the new values of $d$, $K_{t}$, $\kappa_{t}$, and $K_{r}$ as $d_{2}$, $K_{t,2}$, $\kappa_{t,2}$, and $K_{r,2}$, respectively.
						\STATE $D := d + d_{2}$
					\ENDIF
				\ENDIF
			\ENDFOR
		\ENDFOR
		\RETURN $(d,K_{t},\kappa_{t},K_{r})$ and $(d_{2},K_{t,2},\kappa_{t,2},K_{r,2})$ (if any) corresponding to the largest $D$.
	\end{algorithmic}
	\vspace{\baselineskip}
	\hrule
\end{table}

\subsection{Application of the proposed scheme in the uplink channel}

As the achievable DoF of the proposed scheme in the uplink channel is determined by the method given in the previous subsection, it is straightforward that the application of the proposed scheme has to follow the ideas used to develop the method. To simplify the explanation, we present the basic steps of the application in Table \ref{tab:proposed_IA_scheme_application}.

\begin{table}[h]
	\centering
	\caption{Application of the proposed IA scheme in the uplink channel given in \eqref{eq:system_model}}
	\label{tab:proposed_IA_scheme_application}
	\hrule
	\vspace{\baselineskip}
	\begin{algorithmic}[1]
		\STATE Run the algorithm in Table \ref{tab:dof_enhancement} to obtain $(d, K_{t}, \kappa_{t}, K_{r})$ and $(d_{2}, K_{t,2}, \kappa_{t,2}, K_{r,2})$ corresponding to the largest achievable DoF.	
		\FOR{$l := 1$ to $L$} \label{alg:application_for_begin}	
			\STATE Use $(d,K_{t},\kappa_{t},K_{r})$ to set up a sufficient number of IA systems at cell $l$ such that at least $d$ transmit directions can be found by solving the IA systems, and the $K$ users in cell $l$ cause the same amount of ICI to each of $L$ BSs.
			\STATE Solve the IA systems to obtain the precoders of $K$ users in cell $l$.
		\ENDFOR \label{alg:application_for_end}
		\IF{there exists $(d_{2},K_{t,2},\kappa_{t,2},K_{r,2})$}
			\STATE Form the effective channel matrices by right-multiplying the channel matrices by the matrices orthogonal to the already-designed precoders obtained by steps \ref{alg:application_for_begin}--\ref{alg:application_for_end}.
			\STATE Run steps \ref{alg:application_for_begin}--\ref{alg:application_for_end}, but replace $(d,K_{t},\kappa_{t},K_{r})$ with $(d_{2},K_{t,2},\kappa_{t,2},K_{r,2})$.
		\ENDIF		
	\end{algorithmic}
	\vspace{\baselineskip}
	\hrule
\end{table}

\textit{Remark:} It can be seen from \eqref{eq:achievable_dof} that $d$ might be non-integer, as the first term of the min function is not necessarily an integer value. This problem can be resolved by combining the proposed IA scheme with the technique of symbol extension. The details of this combination are omitted here.

\subsection{Upper bound of the total DoF achieved by the proposed scheme}

For simplicity, the upper bound is derived only for the case where the best values of $K_{t}$ and $K_{r}$ are $K$ and $(L-1)$, respectively; however, as will be seen in Section \ref{sec:numerical_analysis}, the total DoF achieved by the proposed scheme is in fact restricted by this upper bound in all cases. Furthermore, it will also be shown that the total DoF achieved by the proposed scheme approaches the upper bound as $K$ increases.

According to \eqref{eq:achievable_dof}, the total achievable DoF of the proposed scheme is not larger than $L K M_{r}/\big( K + (L-1)\kappa_{t} \big)$, and as the constraint \eqref{eq:constraint_kappa} yields 
\[
\kappa_{t} > \frac{ K \big( (L-1)M_{r} - N_{t} \big) }{ (L-1) M_{r} },
\]
the upper bound of the total DoF achieved by the proposed scheme in the uplink channel is given by
\[
D^{\text{UB}} = \frac{ LM_{r}^{2} }{ LM_{r} - N_{t} }.
\]

\section{Numerical Analysis} \label{sec:numerical_analysis}

This section compares the DoF performance of the proposed scheme and the other related IA schemes in the uplink channel of \eqref{eq:system_model}. Through the comparisons, we show that in certain scenarios the proposed scheme can achieve the optimal DoF, and that in other scenarios where it cannot achieve the optimal DoF, it still provides a high DoF gain over the other related IA schemes.

\subsection{Comparisons between DoF performance of the proposed scheme and the optimal DoF}

We compare the optimal total DoF achieved by the structured IA scheme given in \cite{dof_decomposition} and the bounds of the total DoF given in \cite{genie_chain}. The structured IA scheme is applicable only to certain scenarios where $L=2$, whereas the bounds given in \cite{genie_chain} can be used for general multicell multiuser cellular networks. Specifically, when $L = 2,\ K \geq 4$ or $L \geq 3$, and $D_{\text{op}}$ denotes the optimal total DoF, then $D_{\text{op}}$ satisfies \cite{genie_chain}
\[
\left\lbrace
	\begin{array}{l}
		D_{\text{op}} = D_{\text{decom}} = LK \frac{M_{r} N_{t}}{M_{r} + N_{t}}, \text{ if } \frac{M_{r}}{N_{t}} \in \text{ Region 1}, \\
		D_{\text{decom}} \leq D_{\text{op}} \leq D_{\text{proper}} = LK \frac{M_{r} + N_{t}}{LK + 1}, \text{ if } \frac{M_{r}}{N_{t}} \in \text{ Region 2},
	\end{array}
\right.
\]
where $\text{Region 1} = (C^{B}_{\infty},C^{A}_{\infty})$, $\text{Region 2} = (0,C^{B}_{\infty}] \cup [C^{A}_{\infty},\infty)$, $C^{B}_{\infty} = \big( (L-1)K - \sqrt{(L-1)^{2}K^{2} - 4K} \big)/2$, and $C^{A}_{\infty} = \big( (L-1)K + \sqrt{(L-1)^{2}K^{2} - 4K} \big)/2$. Regions 1 and 2 can also be identified on the basis of $D_{\text{proper}}$ and $D_{\text{decom}}$, as $D_{\text{proper}} < D_{\text{decom}}$ in Region 1 and $D_{\text{proper}} \geq D_{\text{decom}}$ in Region 2 \cite{genie_chain}.

\begin{figure}[h]
	\centering
	\includegraphics[trim=80 50 80 60, clip=true, scale=0.36]{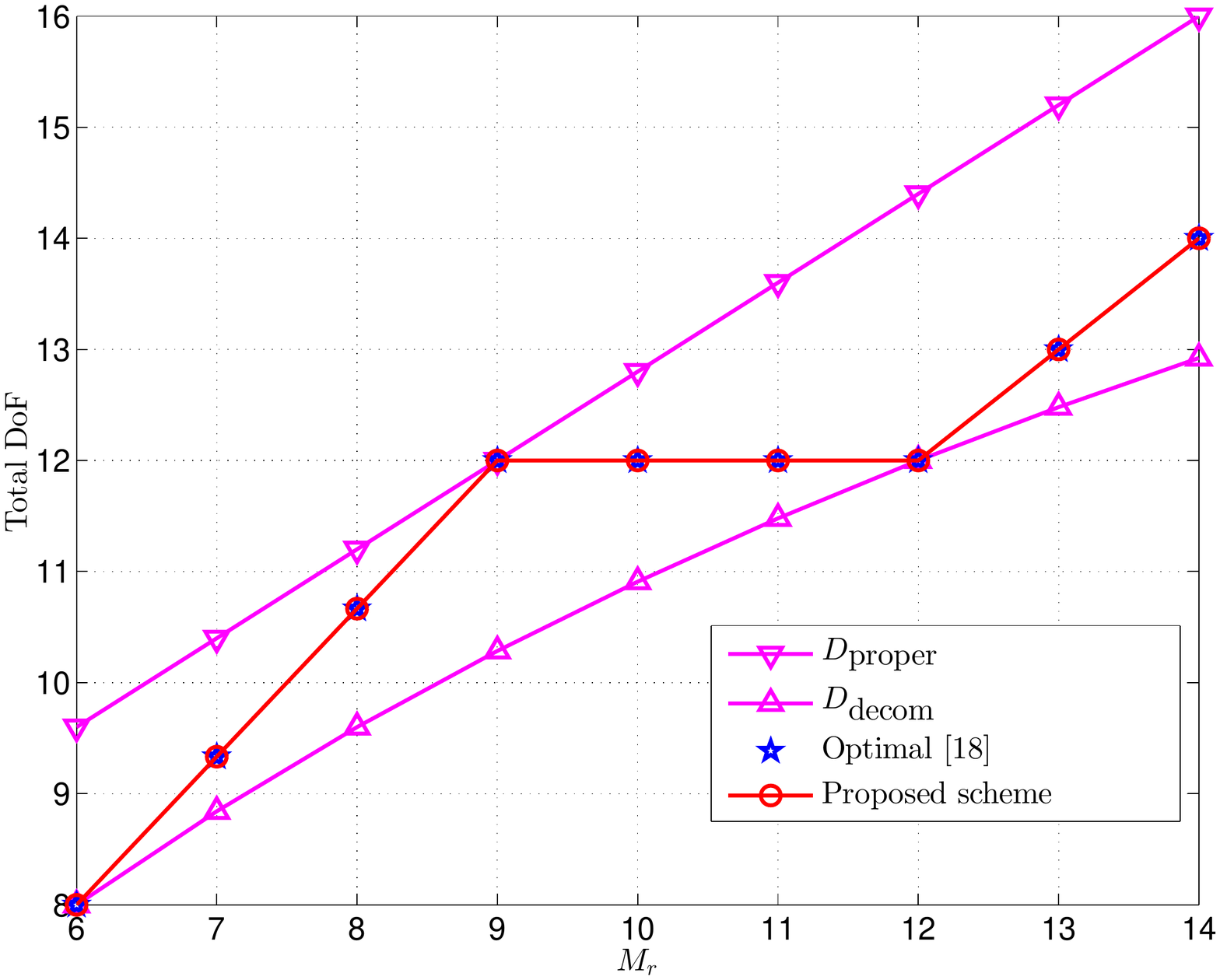}
	\caption{Total DoF versus the number of receive antennas $M_{r}$ when $L = 2$, $K = 2$, and $N_{t} = 6$.}
	\label{fig:proposed_scheme_vs_optimal_Mruns6_14_L2K2N6}
\end{figure}

\begin{figure}[h]
	\centering
	\includegraphics[trim=80 50 80 60, clip=true, scale=0.36]{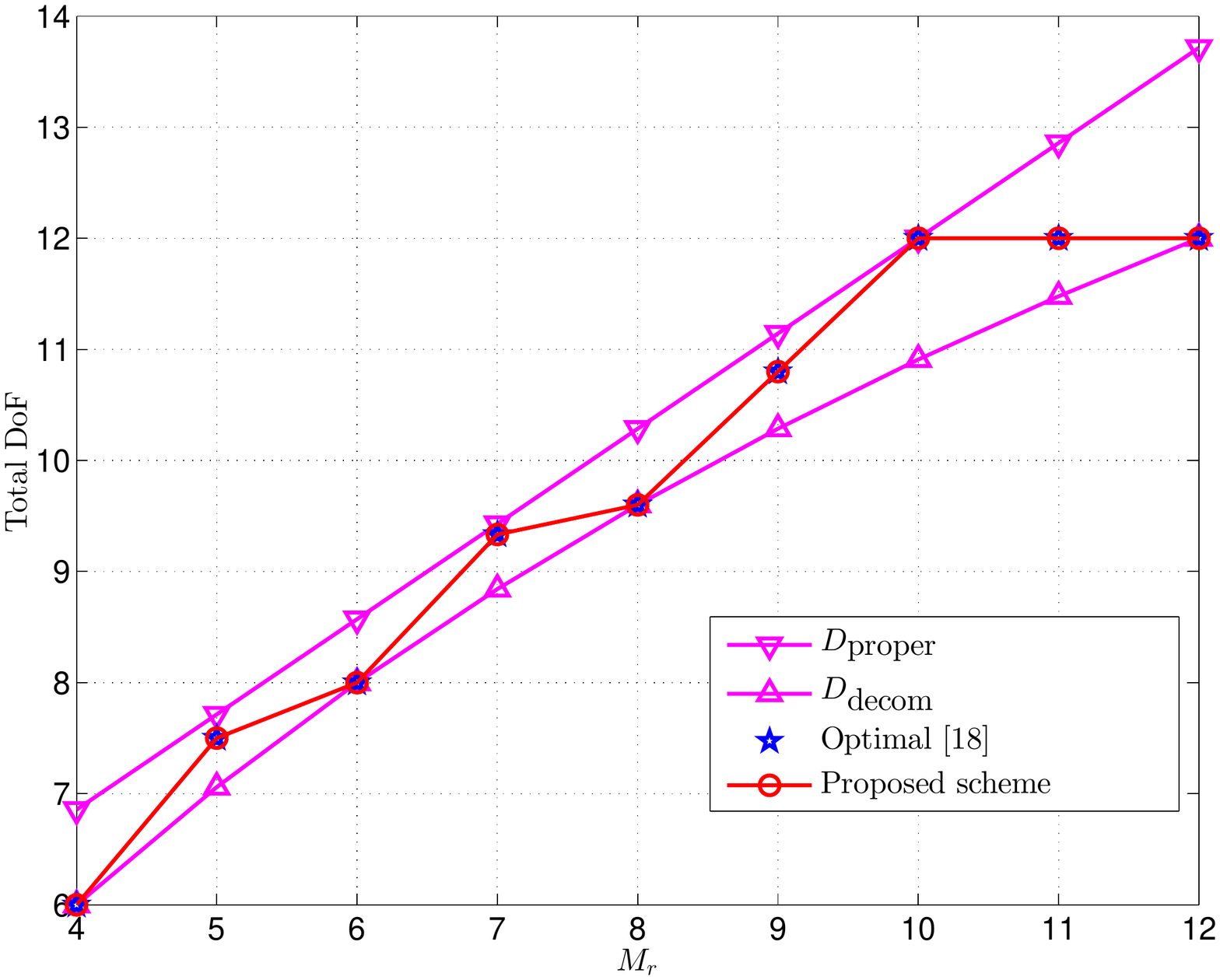}
	\caption{Total DoF versus the number of receive antennas $M_{r}$ when $L = 2$, $K = 3$, and $N_{t} = 4$.}
	\label{fig:proposed_scheme_vs_optimal_Mruns4_12_L2K3N4}
\end{figure}

\begin{figure}[h]
	\centering
	\includegraphics[trim=80 50 80 60, clip=true, scale=0.36]{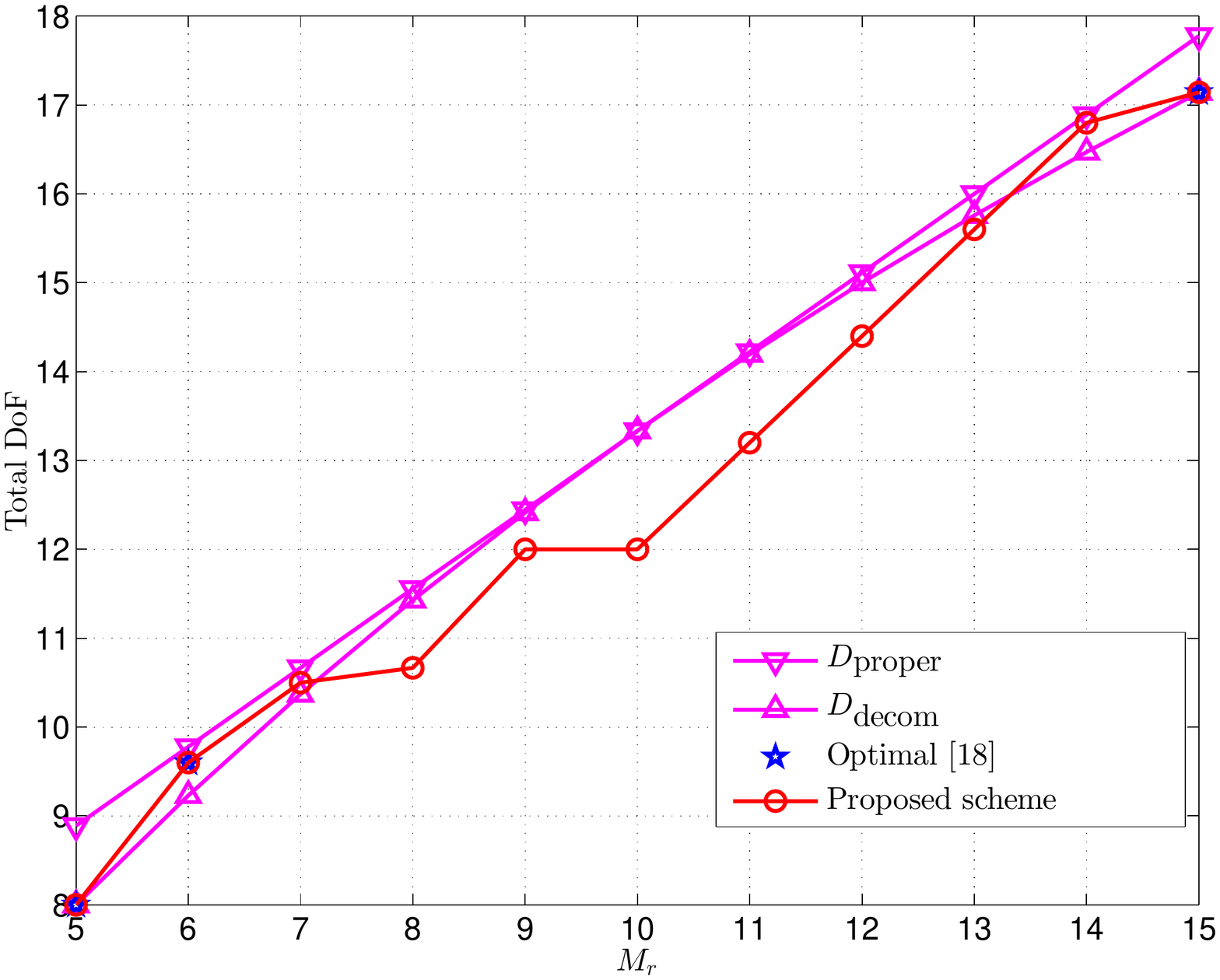}
	\caption{Total DoF versus the number of receive antennas $M_{r}$ when $L = 2$, $K = 4$, and $N_{t} = 5$.}
	\label{fig:proposed_scheme_vs_optimal_Mruns5_15_L2K4N5}
\end{figure}

\begin{figure}[h]
	\centering
	\includegraphics[trim=80 50 80 60, clip=true, scale=0.36]{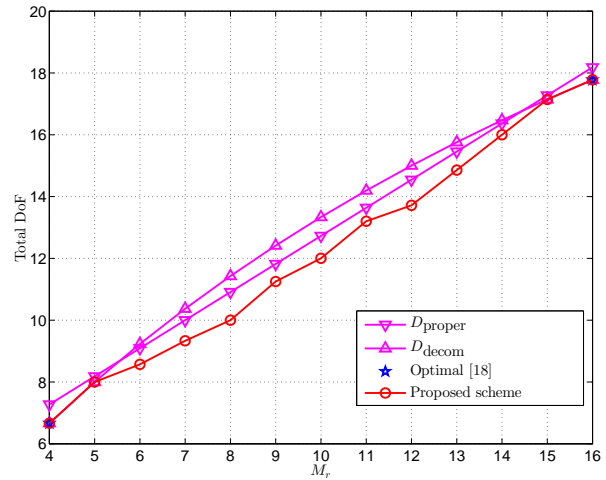}
	\caption{Total DoF versus the number of receive antennas $M_{r}$ when $L = 2$, $K = 5$, and $N_{t} = 4$.}
	\label{fig:proposed_scheme_vs_optimal_Mruns4_16_L2K5N4}
\end{figure}

Figs. \ref{fig:proposed_scheme_vs_optimal_Mruns6_14_L2K2N6}--\ref{fig:proposed_scheme_vs_optimal_Mruns4_16_L2K5N4} present the values of $D_{\text{proper}}$, $D_{\text{decom}}$, the total DoF achieved by the proposed scheme, and that realized by the structured IA scheme for various scenarios where $L = 2$. It can be seen from the figures that the proposed scheme can achieve the optimal DoF in all the cases where the structured IA scheme achieves the optimal DoF. The structured IA scheme is specifically designed for small networks where $L = 2,\ K \in \{2,3\}$ and is developed from the concept of the packing ratio, which is similar to the ICI reduction ratio of $K_{t}/\kappa_{t}$ in this paper but is found manually in \cite{dof_decomposition}; on the other hand, the proposed scheme can be applied to a general multicell multiuser cellular network and can find the best packing ratios automatically. Thus, the proposed scheme can be viewed as a generalization of the structured IA scheme.

Another observation from Figs. \ref{fig:proposed_scheme_vs_optimal_Mruns6_14_L2K2N6}--\ref{fig:proposed_scheme_vs_optimal_Mruns4_16_L2K5N4} is that when $D_{\text{proper}}$ is much larger than $D_{\text{decom}}$, the total DoF achieved by the proposed scheme tends to lie between $D_{\text{proper}}$ and $D_{\text{decom}}$. For example, when $M_{r} \in \{ 5,6,7 \} \cup \{ 14,15 \}$ in Fig. \ref{fig:proposed_scheme_vs_optimal_Mruns5_15_L2K4N5} or when $M_{r} \in \{ 4,5 \} \cup \{ 15,16 \}$ in Fig. \ref{fig:proposed_scheme_vs_optimal_Mruns4_16_L2K5N4}, $D_{\text{proper}}$ is much larger than $D_{\text{decom}}$, and the total DoF of the proposed scheme lies between $D_{\text{proper}}$ and $D_{\text{decom}}$. Because the optimal total DoF $D_{\text{op}}$ also lies between $D_{\text{proper}}$ and $D_{\text{decom}}$ when $D_{\text{proper}} > D_{\text{decom}}$, it can be concluded that when $D_{\text{proper}}$ is much larger than $D_{\text{decom}}$, the DoF achieved by the proposed scheme is either equal to or very close to the optimal DoF. Furthermore, when the gap between $D_{\text{proper}}$ and $D_{\text{decom}}$ is small, even though there are cases where the proposed scheme cannot achieve the optimal DoF, it can still achieve a significant portion of the optimal DoF in these cases, as illustrated in Figs. \ref{fig:proposed_scheme_vs_optimal_Mruns5_15_L2K4N5}--\ref{fig:proposed_scheme_vs_optimal_Mruns4_16_L2K5N4}. Therefore, we conclude that when $D_{\text{proper}}$ is larger than or close to $D_{\text{decom}}$, the proposed scheme can achieve either the optimal DoF or a significant portion of the optimal DoF.

When $D_{\text{decom}}$ is much larger than $D_{\text{proper}}$, the optimal total DoF is given by $D_{\text{op}} = D_{\text{decom}}$; on the other hand, as will be shown in Figs. \ref{fig:proposed_scheme_vs_other_schemes_Kruns2_16_M3N3}--\ref{fig:proposed_scheme_vs_other_schemes_Kruns2_16_M6N4}, the total DoF achieved by the proposed scheme approaches the upper bound $D^{\text{UB}}$ as $K$ increases. Therefore, when $D_{\text{decom}}$ is much larger than $D_{\text{proper}}$ and as $K$ increases, the ratio of the optimal DoF to the DoF of the proposed scheme approaches $\frac{KN_{t}(LM_{r} - N_{t})}{M_{r}(M_{r} + N_{t})}$, from which it can be concluded that the gap between the optimal DoF and the DoF achieved by the proposed scheme increases as the network becomes larger.

\subsection{Comparisons between the proposed scheme and the other related IA schemes}

We compare the coordinated orthogonal scheme (COS), the unstructured IA scheme in \cite{dof_decomposition}, and the IA schemes in \cite{Lee,on_feasibility_l_cell}. To guarantee fair comparisons, symbol extension is incorporated in the schemes given in \cite{Lee,on_feasibility_l_cell}.\footnote{Because it is not trivial to incorporate symbol extension in the scheme in \cite{mu_two_way_relay}, this scheme is not included in the comparisons.} The total DoF achieved by the COS and the schemes in \cite{Lee,on_feasibility_l_cell} are presented in closed forms in Table \ref{tab:total_dofs}, whereas the total DoF achieved by the scheme in \cite{dof_decomposition} cannot be presented in closed form and is obtained numerically. Owing to the similarity between the proposed scheme and the unstructured IA scheme, we first compare only these two algorithms, and then we compare the proposed scheme, the COS, and the schemes in \cite{Lee,on_feasibility_l_cell}.

\begin{table}[h]
	\centering
	\caption{Total DoF achieved by the other IA schemes}
	\label{tab:total_dofs}
	\renewcommand{\arraystretch}{1.4}
	\begin{tabular}{|c|l|}
		\hline
		& \multicolumn{1}{c|}{Total achievable DoF} \\
		\hline
		COS & $\min(M_r,LKN_t).$ \\
		\hline
		\cite{Lee} ($L=2$) & $\left\lbrace
		\begin{array}{l}
		2N_{t}, \text{ if }N_t \leq KM/(K+1), \\
		2KM/(K+1),\text{ otherwise.}
		\end{array} \right.$ \\
		\hline		
		\cite{on_feasibility_l_cell} & $LK\min \big(  \frac{M_{r}}{LK-1} , L(KN_{t} - M_{r}) \big)$. \\
		\hline
	\end{tabular}
	\centering
\end{table}

\begin{figure}[h]
	\centering
	\includegraphics[trim= 5 0 5 5, clip=true, scale=0.32]{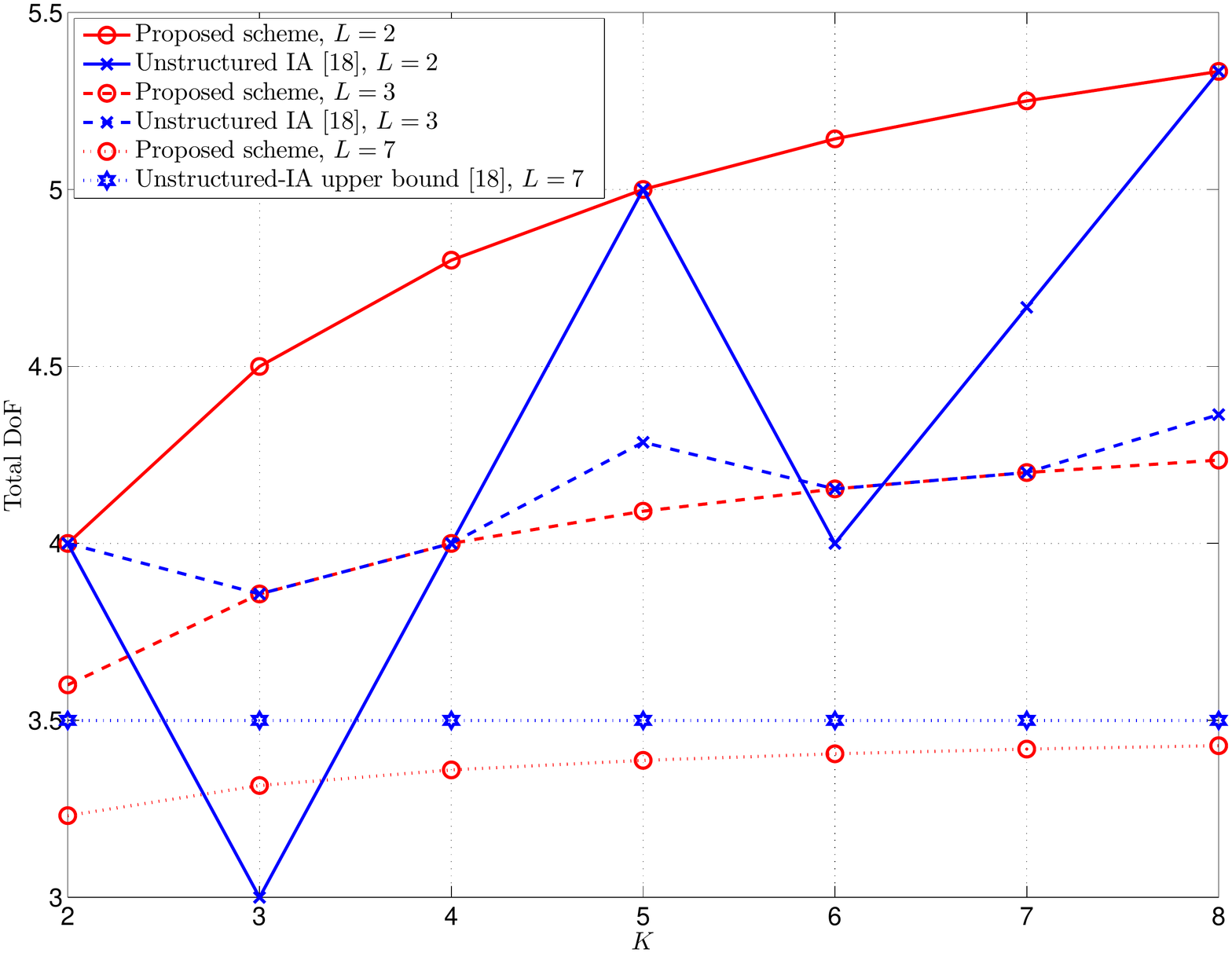}
	\caption{Total DoF versus the number of users in each cell $K$ when $M_{r} = 3$ and $N_{t} = 3$.}
	\label{fig:proposed_scheme_vs_unstructured_scheme}
\end{figure}

Fig. \ref{fig:proposed_scheme_vs_unstructured_scheme} presents the total DoF achieved by the proposed scheme and that realized by the unstructured IA scheme for various scenarios. To guarantee fair comparisons, the maximum length of symbol extension used by the unstructured scheme is set to be equal to the length of symbol extension used by the proposed scheme. From the figures, it is observed that when $L = 2$, the proposed scheme outperforms the unstructured scheme and that when $L = 3$, the proposed scheme is slightly outperformed by the unstructured scheme. Because the unstructured scheme can obtain an ICI reduction ratio of $(L-1) K d/ \tau$, where $\tau < (L-1) K d$, whereas the proposed scheme can obtain only the ratio $K_{t}/\kappa_{t}$, the unstructured scheme has more options for selecting the reduction ratio when $L > 2$; hence, the unstructured scheme can achieve a better IA than the proposed scheme. However, when the number of users in each cell increases, the variety in the reduction ratio of the proposed scheme also increases, giving the scheme more options for performing IA, and as can be seen in the figure for $L = 3$, the gap between the two schemes tends to decrease when $K$ increases.

Fig. \ref{fig:proposed_scheme_vs_unstructured_scheme} also shows that when $L = 7$ and as $K$ increases, the total DoF achieved by the proposed scheme approaches the upper bound of the total DoF achieved by the unstructured scheme, which is given by $L M_{r}^{2}/(L M_{r} - N_{t})$ \cite{dof_decomposition}. Because this upper bound is strictly larger than the total DoF achieved by the unstructured scheme, it can be concluded that, when the network is large, i.e., when both $L$ and $K$ are large, the total DoF of the proposed scheme approaches that of the unstructured scheme.

Even though the unstructured IA scheme can in some cases achieve more DoF than the proposed scheme, the latter still provides the following two advantages over the former. First, the achievable DoF of the proposed scheme is analytically tractable, but that of the unstructured scheme is intractable. Second, the proposed scheme requires a considerably lower complexity than the unstructured scheme. To find the precoders for all users in the channel, the proposed scheme in the worst case needs to solve $L \binom{K}{K_{t}}$ IA systems, each of which requires about $\mathcal{O}\big( (K_{t}SN)^{3} \big)$ complex operations, whereas the unstructured IA scheme needs to solve one IA system, which requires about $\mathcal{O}\big( (LKS^{2}dN)^{3} \big)$ complex operations, where $S$ is the length of the symbol extension. Furthermore, because the $L \binom{K}{K_{t}}$ IA systems that the proposed scheme needs to solve are independent of each other, the actual complexity of the proposed scheme can be significantly reduced by employing parallel computing; in contrast, the actual complexity of the unstructured scheme is as manifold as $\mathcal{O}( (LKS^{2}dN)^{3} )$ because the unstructured scheme does not guarantee the decodability of the desired signals; thus, it needs to be executed many times, each of which corresponds to a different number of desired signals, until the decodability is verified.

\begin{figure}[h]
	\centering
	\includegraphics[trim= 5 0 5 5, clip=true, scale=0.33]{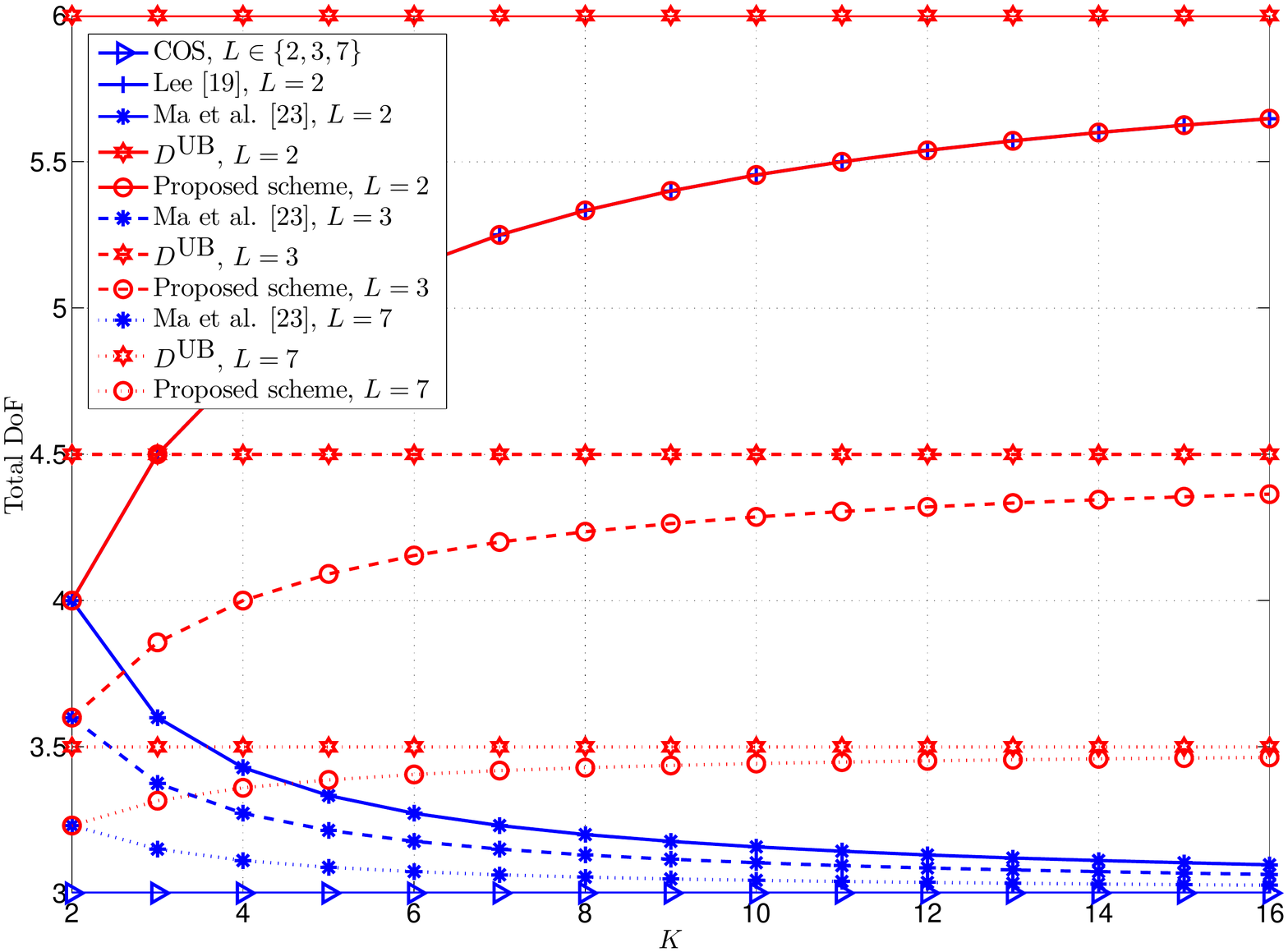}
	\caption{Total DoF versus the number of users in each cell $K$ when $M_{r} = 3$ and $N_{t} = 3$.}
	\label{fig:proposed_scheme_vs_other_schemes_Kruns2_16_M3N3}
\end{figure}

\begin{figure}[h]
	\centering
	\includegraphics[trim= 5 0 5 5, clip=true, scale=0.33]{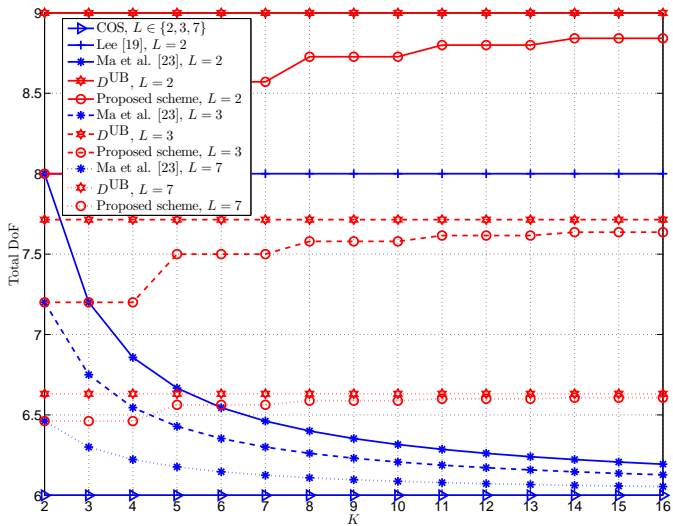}
	\caption{Total DoF versus the number of users in each cell $K$ when $M_{r} = 6$ and $N_{t} = 4$.}
	\label{fig:proposed_scheme_vs_other_schemes_Kruns2_16_M6N4}
\end{figure}

Figs. \ref{fig:proposed_scheme_vs_other_schemes_Kruns2_16_M3N3}--\ref{fig:proposed_scheme_vs_other_schemes_Kruns2_16_M6N4} present the total DoF achieved by the proposed scheme and its upper bound $D^{\text{UB}}$, the total DoF achieved by the COS, and those realized by the schemes in \cite{Lee,on_feasibility_l_cell} for various scenarios. The figures show that the proposed scheme outperforms the other schemes in all cases. The DoF gain of the proposed scheme over the other schemes is significant when $L$ is small, but it tends to decrease as $L$ increases. This is because as $L$ increases, the DoF performance of both the proposed scheme and the other schemes suffers from the bottleneck caused by the intensification of the ICI.

Despite the reductions in the gaps between the proposed scheme and the other schemes as $L$ increases, the proposed scheme tends to provide a constant gain over the COS and the scheme in \cite{on_feasibility_l_cell} as $K$ increases. Because the total DoF achieved by the scheme in \cite{on_feasibility_l_cell} is given by $LKM_{r}/(LK - 1)$, as $K$ increases it approaches $M_{r}$, which is also the total DoF achieved by the COS. On the other hand, as can be seen in Figs. \ref{fig:proposed_scheme_vs_other_schemes_Kruns2_16_M3N3}--\ref{fig:proposed_scheme_vs_other_schemes_Kruns2_16_M6N4}, as $K$ increases, the total DoF achieved by the proposed scheme approaches its upper bound, which is given by $LM_{r}^{2}/(LM_{r} - N_{t})$. Therefore, as $K$ increases, the total DoF achieved by the proposed scheme is approximately $LM_{r}/(LM_{r} - N_{t})$ times as many as those achieved by the COS and the scheme in \cite{on_feasibility_l_cell}; in particular, when $M_{r} = N_{t}$, the DoF gain over the COS is simplified to $L/(L-1)$.

\section{Conclusion} \label{sec:conclusion}

A new signal-space linear non-iterative IA scheme that can be applied to a general uplink channel is proposed in this paper. The proposed scheme can reduce the ICI by a ratio of $K_{t}/\kappa_{t}$, as it can consolidate the ICI from $K_{t}$ users into a subspace that is as large as a subspace created by the ICI from $\kappa_{t}$ users, where $K_{t}$ and $\kappa_{t}$ are chosen so that the ratio is the largest one that can guarantee the decodability of the transmitted signals. Consequently, numerical analysis shows that in certain scenarios the proposed scheme can achieve the optimal DoF, and in other scenarios where the proposed scheme cannot achieve the optimal DoF, it can still provide DoF gains over the other signal-space linear non-iterative IA schemes. Furthermore, as the number of users in each cell increases, the proposed scheme provides a constant DoF gain of $LM_{r}/(LM_{r} - N_{t})$ over the conventional coordinated orthogonal scheme.

\appendices

\section{Proof of Lemma \ref{lemma_randomized_sum}}  \label{proof_lemma_randomized_sum}

For $n_{2} = n_{1}$, \textit{Lemma \ref{lemma_randomized_sum}} is correct as we always have
\[
 \text{dim}\Big( \mathcal{C} \big( \mathbf{A}_{1},\mathbf{A}_{2},\cdots,\mathbf{A}_{n_{1}} \big) \Big) \leq \sum_{k=1}^{n_{1}}\text{dim}\Big( \mathcal{C} \big( \mathbf{A}_{k} \big) \Big) \leq n_{1}r.
\]

For $0 \leq n_{2} < n_{1}$ and $q = n_{1} - n_{2}$, \eqref{eq:lemma3} can be rewritten as
\begin{equation} \label{lemma3_system_of_eqns}
 \left\lbrace
   \begin{aligned}
     \gamma_{11}\mathbf{A}_1 + \cdots + \gamma_{1q}\mathbf{A}_{q} & = -\gamma_{1(q+1)}\mathbf{A}_{q+1} - \cdots - \gamma_{1n_{1}}\mathbf{A}_{n_{1}}, \\
     \gamma_{21}\mathbf{A}_1 + \cdots + \gamma_{2q}\mathbf{A}_{q} & = -\gamma_{2(q+1)}\mathbf{A}_{q+1} - \cdots - \gamma_{2n_{1}}\mathbf{A}_{n_{1}}, \\
     & \vdots \\
     \gamma_{q1}\mathbf{A}_1 + \cdots + \gamma_{qq}\mathbf{A}_{q} & = -\gamma_{q(q+1)}\mathbf{A}_{q+1} - \cdots - \gamma_{qn_{1}}\mathbf{A}_{n_{1}}.
   \end{aligned}
 \right.
\end{equation}
If we consider \eqref{lemma3_system_of_eqns} as a system of linear equations with respect to the \textquotedblleft variables\textquotedblright \ $\mathbf{A}_{k}, \ k=1,2,\cdots,q$, then the coefficient matrix of (\ref{lemma3_system_of_eqns}) is given by
\[
\mathbf{\Gamma} = \begin{bmatrix}
 \gamma_{11} & \gamma_{12} & \cdots & \gamma_{1q} \\
 \gamma_{21} & \gamma_{22} & \cdots & \gamma_{2q} \\
 \vdots & \vdots & \ddots & \vdots \\
 \gamma_{q1} & \gamma_{q2} & \cdots & \gamma_{qq} \\
\end{bmatrix}.
\]
As $\gamma_{ik}, \ i = 1,2,\cdots,q; \ k=1,2,\cdots,n_{1}$ are i.i.d. scalars, $\mathbf{\Gamma}$ is almost surely of full rank. Thus, each matrix $\mathbf{A}_k, \ k=1,2,\cdots,q$ can be expressed as a linear combination of $(n_{1} - q)$ matrices $\mathbf{A}_{k}, \ k = q+1, q+2,\cdots,n_{1}$. It follows that
\[
\mathcal{C} (\mathbf{A}_1, \mathbf{A}_2, \cdots, \mathbf{A}_{n_{1}}) \equiv \mathcal{C} (\mathbf{A}_{n_{1}-n_{2}+1}, \mathbf{A}_{n_{1}-n_{2}+2}, \cdots, \mathbf{A}_{n_{1}}).
\]
Therefore,
\begin{multline*}
\text{dim}\Big( \mathcal{C} \big( \mathbf{A}_1, \ \mathbf{A}_2, \ \cdots, \ \mathbf{A}_{n_{1}} \big) \Big) \leq \\
\sum_{k=n_{1}-n_{2}+1}^{n_{1}} \text{dim}\Big( \mathcal{C} \big( \mathbf{A}_{k} \big) \Big) \leq n_{2} r.
\end{multline*}
The equality occurs when $n_{2}$ subspaces $\mathcal{C} (\mathbf{A}_{k}), \ k = n_{1}-n_{2}+1, n_{1}-n_{2}+2,\cdots, n_{1}$ are of dimension $r$ and disjoint and the containing space is large enough, i.e., $m_{1} \geq n_{2} r$.

\section{} \label{app:rank_coef_matrix}

\begin{lemma}
\label{lemma:rank_coef_matrix}
Given the coefficient matrix $\mathbf{C}$ in \eqref{eq:design_precoders_for_users_in_one_cell_extended}, if $N_{t} \leq K_{r} M_{r}$, then $\mathbf{C}$ is almost surely a full rank matrix, i.e., $\text{rank}\big( \mathbf{C} \big) = \min\big( K_{r} ( K_{t} - \kappa_{t} ) M_{r}, K_{t} N_{t} \big)$, for any $K_{t} \geq 1$.
\end{lemma}
\begin{IEEEproof}
	
The basic idea of the proof is to show that $\text{Pr}\big( \text{det}( \mathbf{C} ) = 0 \big) = 0$, where Pr() and det() denote the probability and determinant functions, respectively, by iteratively using row operations and cofactor expansions along columns. The row operations aim to zero out some entries in the target matrix in order to produce a resultant matrix to which we can apply the main argument of the proof, which is based on cofactor expansions along columns and is specifically used to reach the conclusion that
\[
\text{Pr}\big( \text{det}( \mathbf{A}_{\bar{1}} ) = 0 \big) = 0 \Rightarrow \text{Pr}\big( \text{det}( \mathbf{A} ) = 0 \big) = 0,
\]
where $\mathbf{A}$ is some given square matrix, and $\mathbf{A}_{\bar{n}}$ is the matrix obtained by stripping both the first $n$ columns and the first $n$ rows from $\mathbf{A}$.

Before presenting the proof in detail, we give a simple example to illustrate the basic idea of the proof. In this example, we assume that $K_{r} = 2$, $K_{t} = 4$, $\kappa_{t} = 2$, and $M_{r} = N_{t}$. Furthermore, for the sake of a simple representation, we use the symbol \textquotedblleft$\times$\textquotedblright \ to represent any scalar number. The matrix $\mathbf{C}$ for this example is then given by
\[
\mathbf{C} =
\begin{bmatrix}
	\times \mathbf{H}_{1 1} & \times \mathbf{H}_{1 2} & \times \mathbf{H}_{1 3} & \times \mathbf{H}_{1 4} \\
	\times \mathbf{H}_{1 1} & \times \mathbf{H}_{1 2} & \times \mathbf{H}_{1 3} & \times \mathbf{H}_{1 4} \\
	\times \mathbf{H}_{2 1} & \times \mathbf{H}_{2 2} & \times \mathbf{H}_{2 3} & \times \mathbf{H}_{2 4} \\
	\times \mathbf{H}_{2 1} & \times \mathbf{H}_{2 2} & \times \mathbf{H}_{2 3} & \times \mathbf{H}_{2 4} \\
\end{bmatrix}.
\]

The first step in the proof is to use row operations to transform $\mathbf{C}$ into $\tilde{\mathbf{C}}$, which is given by
\[
\tilde{\mathbf{C}} =
\begin{bmatrix}
\times \mathbf{H}_{1 1} & \times \mathbf{H}_{1 2} & \times \mathbf{H}_{1 3} & \times \mathbf{H}_{1 4} \\
\times \mathbf{H}_{2 1} & \times \mathbf{H}_{2 2} & \times \mathbf{H}_{2 3} & \times \mathbf{H}_{2 4} \\
\mathbf{0} & \times \mathbf{H}_{1 2} & \times \mathbf{H}_{1 3} & \times \mathbf{H}_{1 4} \\
\mathbf{0} & \times \mathbf{H}_{2 2} & \times \mathbf{H}_{2 3} & \times \mathbf{H}_{2 4} \\
\end{bmatrix}.
\]
By applying cofactor expansion along the first column of $\tilde{\mathbf{C}}$, we have
\[
\text{det}( \tilde{\mathbf{C}} ) = \sum_{k \in \{ 1, 2 \}} \sum_{i = 1}^{M_{r}} h_{k 1}^{i 1} \tilde{C}_{k 1}^{i 1},
\]
where $h_{k l}^{i j}$ denotes the $(i,j)$ entry in $\mathbf{H}_{kl}$, and $\tilde{C}_{k l}^{i j}$ is the cofactor corresponding to the position of $h_{k l}^{i j}$ in $\tilde{\mathbf{C}}$. Then the main argument, which will be described later, is used $N_{t}$ times to yield
\begin{multline*}
\text{Pr}\big( \text{det}( \tilde{\mathbf{C}}_{\overline{N_{t}}} ) = 0 \big) = 0 \Rightarrow \text{Pr}\big( \text{det}( \tilde{\mathbf{C}}_{\overline{(N_{t} - 1)}} ) = 0 \big) = 0 \Rightarrow \\
\cdots \Rightarrow \text{Pr}\big( \text{det}( \tilde{\mathbf{C}} ) = 0 \big) = 0.
\end{multline*}
Because the form of
\[
\tilde{\mathbf{C}}_{\overline{N_{t}}} = \begin{bmatrix}
\times \mathbf{H}_{2 2} & \times \mathbf{H}_{2 3} & \times \mathbf{H}_{2 4} \\
\times \mathbf{H}_{1 2} & \times \mathbf{H}_{1 3} & \times \mathbf{H}_{1 4} \\
\times \mathbf{H}_{2 2} & \times \mathbf{H}_{2 3} & \times \mathbf{H}_{2 4} \\
\end{bmatrix}
\]
is similar to that of $\mathbf{C}$, the entire process of applying row operations and the main argument can be iteratively applied to $\tilde{\mathbf{C}}_{\overline{N_{t}}}$ to yield
\begin{multline*}
\text{Pr}\big( \text{det}( \tilde{\mathbf{C}}_{\overline{2N_{t}}} ) = 0 \big) = 0 \Rightarrow \text{Pr}\big( \text{det}( \tilde{\mathbf{C}}_{\overline{(2N_{t} - 1)}} ) = 0 \big) = 0 \Rightarrow \\
\cdots \Rightarrow \text{Pr}\big( \text{det}( \tilde{\mathbf{C}}_{\overline{N_{t}}} ) = 0 \big) = 0.
\end{multline*}
Similarly, the entire process can also be applied to $\tilde{\mathbf{C}}_{\overline{2N_{t}}}$ to yield
\begin{multline*}
\text{Pr}\big( \text{det}( \tilde{\mathbf{C}}_{\overline{3N_{t}}} ) = 0 \big) = 0 \Rightarrow \text{Pr}\big( \text{det}( \tilde{\mathbf{C}}_{\overline{(3N_{t} - 1)}} ) = 0 \big) = 0 \Rightarrow \\
\cdots \Rightarrow \text{Pr}\big( \text{det}( \tilde{\mathbf{C}}_{\overline{2N_{t}}} ) = 0 \big) = 0.
\end{multline*}
As $\tilde{\mathbf{C}}_{\overline{3N_{t}}} = \times \mathbf{H}_{2 4}$ is almost surely of full column rank, it is almost surely that $\text{Pr}\big( \text{det}( \tilde{\mathbf{C}}_{\overline{3N_{t}}} ) = 0 \big) = 0$, which, according to the deduction chain, leads to $\text{Pr}\big( \text{det}( \tilde{\mathbf{C}} ) = 0 \big) = 0$, and hence $\text{Pr}\big( \text{det}( \mathbf{C} ) = 0 \big) = 0$, as $\tilde{\mathbf{C}}$ is obtained from $\mathbf{C}$ by using row operations. This concludes the proof for the given example.

The details of the proof are presented in the following and are divided into two cases, one when $\mathbf{C}$ is a square matrix, and one when $\mathbf{C}$ is not a square matrix.

\textit{Case 1) $\mathbf{C}$ is a square matrix.}

Let us denote $\tilde{\mathbf{C}}$ as the matrix resulting from applying row operations on $\mathbf{C}$, then $\mathbf{C}$ is of full rank if and only if $\tilde{\mathbf{C}}$ is of full rank, as the rank of a matrix is not affected by row operations.

The row operations on $\mathbf{C}$ are chosen so that they will produce $\tilde{\mathbf{C}}$ given by \eqref{eq:c_tilde} at the top of the next page, where $\delta_{ij}^{k} = \frac{ \gamma_{ 11 }^{k} }{ \gamma_{ i1 }^{k} } \gamma_{ij}^{k} - \gamma_{1j}^{k}$. Specifically, the row operations are performed as follows. First, row-exchange operations are used to move the row blocks containing $\gamma_{11}^{k} \mathbf{H}_{k 1},\ \forall k = 1,2,\cdots,K_{r}$ into the top row-block positions. Second, row-addition operations are used to zero out all the entries of $\gamma_{i1}^{k} \mathbf{H}_{k 1},\ \forall i = 2,3,\cdots,K_{t} - \kappa_{t}$, which are located in the first $N_{t}$-column. It can be readily seen that this second step can always be done. Third, row-multiplication operations are used to normalize $\gamma_{11}^{k} \mathbf{H}_{k 1}$ into $\mathbf{H}_{k 1},\ \forall k = 1,2,\cdots,K_{r}$.

\begin{figure*}[t]
	\begin{equation} \label{eq:c_tilde}	
	\tilde{\mathbf{C}} =
	\left\lbrack \begin{array}{cccc}
	\mathbf{H}_{11} & \frac{\gamma_{12}^{1}}{\gamma_{11}^{1}} \mathbf{H}_{12} & \cdots & \frac{\gamma_{1K_{t}}}{\gamma_{11}^{1}} \mathbf{H}_{1K_{t}} \\
	\mathbf{H}_{21} & \frac{\gamma_{12}^{2}}{\gamma_{11}^{2}} \mathbf{H}_{22} & \cdots & \frac{\gamma_{1K_{t}}^{2}}{\gamma_{11}^{2}} \mathbf{H}_{2K_{t}} \\
	\vdots & \vdots & \ddots & \vdots \\
	\mathbf{H}_{K_{r}1} & \frac{\gamma_{12}^{K_{r}}}{\gamma_{11}^{K_{r}}} \mathbf{H}_{K_{r}2} & \cdots & \frac{\gamma_{1K_{t}}^{K_{r}}}{\gamma_{11}^{K_{r}}} \mathbf{H}_{K_{r}K_{t}} \\
	\mathbf{0} & \delta_{22}^{1} \mathbf{H}_{12} & \cdots & \delta_{2K_{t}}^{1} \mathbf{H}_{1K_{t}} \\
	\vdots & \vdots & \ddots & \vdots \\
	\mathbf{0} & \delta_{(K_{t} - \kappa_{t})2}^{1} \mathbf{H}_{12} & \cdots & \delta_{(K_{t} - \kappa_{t})K_{t}}^{1} \mathbf{H}_{1K_{t}} \\
	\mathbf{0} & \delta_{22}^{2} \mathbf{H}_{22} & \cdots & \delta_{2K_{t}}^{2} \mathbf{H}_{2K_{t}} \\
	\vdots & \vdots & \ddots & \vdots \\
	\mathbf{0} & \delta_{(K_{t} - \kappa_{t})2}^{2} \mathbf{H}_{22} & \cdots & \delta_{(K_{t} - \kappa_{t})K_{t}}^{2} \mathbf{H}_{2K_{t}} \\		
	\vdots & \vdots & \ddots & \vdots \\
	\mathbf{0} & \delta_{22}^{K_{r}} \mathbf{H}_{K_{r}2} & \cdots & \delta_{2K_{t}}^{K_{r}} \mathbf{H}_{K_{r}K_{t}} \\
	\vdots & \vdots & \ddots & \vdots \\
	\mathbf{0} & \delta_{(K_{t} - \kappa_{t})2}^{K_{r}} \mathbf{H}_{K_{r}2} & \cdots & \delta_{(K_{t} - \kappa_{t})K_{t}}^{K_{r}} \mathbf{H}_{K_{r}K_{t}}
	\end{array} \right\rbrack.
	\end{equation}
	\hrulefill
	\vspace{4pt}
\end{figure*}

The main argument of the proof can be applied on $\tilde{\mathbf{C}}$ as follows. Without loss of generality, we perform the cofactor expansion along the first column of $\tilde{\mathbf{C}}$ to yield that
\[
\text{det}( \tilde{\mathbf{C}} ) = \sum_{k = 1}^{K_{r}} \sum_{i = 1}^{M_{r}} h_{k 1}^{i 1} \tilde{C}_{k 1}^{i 1},
\]
Thus, $\text{det}\big( \tilde{\mathbf{C}} \big)$ can be considered as a function of the variables $h_{k 1}^{i 1},\ k = 1,2,\cdots,K_{r},\ i = 1,2,\cdots,M_{r}$, and hence $\text{det}\big( \tilde{\mathbf{C}} \big) = 0$ only if one of the following events occurs:
\begin{enumerate}
\item $h_{k 1}^{i 1},\ k = 1,2,\cdots,K_{r},\ i = 1,2,\cdots,M_{r}$ are the roots of the equation $\text{det}\big( \tilde{\mathbf{C}} \big) = 0$.
\item $\tilde{C}_{k 1}^{i 1},\ k = 1,2,\cdots,K_{r},\ i = 1,2,\cdots,M_{r}$ are all zeros.
\end{enumerate}
As $h_{k 1}^{i 1},\ \forall k,i$ are drawn from a continuous distribution and are all independent of $\tilde{C}_{k 1}^{i 1},\ \forall k,i$, the probability that $h_{k 1}^{i 1},\ \forall k,i$ are the roots of $\text{det}\big( \tilde{\mathbf{C}} \big) = 0$ is equal to zero \cite{IA_DoF_X}. In other words, it is almost surely that the first event cannot occur. Thus, the event of $\text{det}\big( \tilde{\mathbf{C}} \big) = 0$ is solely dependent on the second event, and hence we have
\begin{equation} \label{eq:main_argument_proof_lemma_rank_coef_matrix_second_event_1}
\text{Pr}( \text{the second event} ) = 0 \Rightarrow \text{Pr}\big( \text{det}( \tilde{\mathbf{C}} ) = 0 \big) = 0.
\end{equation}
Owing to the requirement of the second event, we can deduce that
\begin{equation} \label{eq:main_argument_proof_lemma_rank_coef_matrix_second_event_2}
\text{Pr}( \tilde{C}_{1 1}^{1 1} = 0 ) = 0 \Rightarrow \text{Pr}( \text{the second event} ) = 0.
\end{equation}
From \eqref{eq:main_argument_proof_lemma_rank_coef_matrix_second_event_1} and \eqref{eq:main_argument_proof_lemma_rank_coef_matrix_second_event_2} and as $\tilde{C}_{1 1}^{1 1}$ is the cofactor of the $(1,1)$ entry in $\tilde{\mathbf{C}}$, we conclude that
\[
\text{Pr}\big( \text{det}( \tilde{\mathbf{C}}_{\bar{1}} ) = 0 \big) = 0 \Rightarrow \text{Pr}\big( \text{det}( \tilde{\mathbf{C}} ) = 0 \big) = 0.
\]

As the first column in $\tilde{\mathbf{C}}_{\bar{1}}$ and that in $\tilde{\mathbf{C}}$ have similar form, i.e., the non-zero entries in each column are i.i.d. and independent of all of the other entries in the matrix, we can also apply the main argument on $\tilde{\mathbf{C}}_{\bar{1}}$ and consequently arrive at the following conclusion:
\[
\text{Pr}\Big( \text{det}\big( \tilde{\mathbf{C}}_{\bar{2}} \big) = 0 \Big) = 0 \Rightarrow \text{Pr}\Big( \text{det}\big( \tilde{\mathbf{C}}_{\bar{1}} \big) = 0 \Big) = 0.
\]
Similarly, the argument can further be used on $\tilde{\mathbf{C}}_{\bar{2}}$, $\tilde{\mathbf{C}}_{\bar{3}}$, $\cdots$, $\tilde{\mathbf{C}}_{N_{t}}$ to yield that
\begin{multline} \label{eq:probability_chain}
\text{Pr}\Big( \text{det}\big( \tilde{\mathbf{C}}_{\overline{N_{t}}} \big) = 0 \Big) = 0 \Rightarrow \text{Pr}\Big( \text{det}\big( \tilde{\mathbf{C}}_{\overline{(N_{t}-1)}} \big) = 0 \Big) = 0 \Rightarrow \\
\cdots \Rightarrow \text{Pr}\Big( \text{det}\big( \tilde{\mathbf{C}} \big) = 0 \Big) = 0.
\end{multline}
It is noted that the argument can be applied $N_{t}$ times only if $N_{t} \leq K_{r} M_{r}$, as otherwise the first column in $\tilde{\mathbf{C}}_{\overline{K_{r} M_{r}}}$ is a zero vector, and the argument cannot be used.

After the main argument is applied $N_{t}$ times, the entire process, including performing row operations and applying the main argument, is then re-applied on $\tilde{\mathbf{C}}_{\overline{N_{t}}}$ in order to arrive at a conclusion similar to \eqref{eq:probability_chain}. This is possible because $\tilde{\mathbf{C}}_{\overline{N_{t}}}$ and the original matrix $\mathbf{C}$ have the same form as can be seen from \eqref{eq:c_tilde} and \eqref{eq:design_precoders_for_users_in_one_cell_extended}. Therefore, we can further conclude that the entire process can be repeated until we reach the last matrix with only one entry.

Without loss of generality, we assume that the row-exchange operations do not move the last row of its target matrix, and hence the entry in the last matrix must be a product between $h_{ K_{r} K_{t} }^{ M_{r} N_{t} }$ and a function of some scalars among $\{ \gamma_{ij}^{k},\ i=1,2,\cdots,K_{t} - \kappa_{t},\ j=1,2,\cdots,K_{t},\  k=1,2,\cdots,K_{r} \}$. As $h_{ K_{r} K_{t} }^{ M_{r} N_{t} }$ and all scalars $\gamma_{ij}^{k}$ are independent of each other and are drawn from continuous distributions, Pr(the entry in the last matrix = 0) = 0, thereby leading to that Pr(the determinant of the last matrix = 0) = 0.

By following the deduction chain whose beginning point is that Pr(the entry in the last matrix = 0) = 0, we arrive at $\text{Pr}\Big( \text{det}\big( \tilde{\mathbf{C}} \big) = 0 \Big) = 0$, which leads to $\text{Pr}\Big( \text{det}\big( \mathbf{C} \big) = 0 \Big) = 0$. Therefore, it can be concluded that $\mathbf{C}$ is almost surely of full rank.

\textit{Case 2) $\mathbf{C}$ is not a square matrix.}

By using the proof given for \textit{Case 1}, we can show that the $\theta \times \theta$ leading principal matrix of $\mathbf{C}$ is almost surely of full rank, where $\theta = \min \big( K_{r}( K_{t} - \kappa_{t} )M_{r}, K_{t}N_{t} \big)$; thus, $\mathbf{C}$ is almost surely of full rank. Therefore, the proof is concluded.
\end{IEEEproof}

\section{Proof that the precoders found by solving \eqref{eq:design_precoders_for_users_in_one_cell_extended} are of full column rank} \label{app:proof_full_rank_precoders}

Let us assume that $\mathbf{V}_{1}$ is not a full column rank matrix, then there exist $d$ scalars $\{ \lambda_{i},\ i=1,2,\cdots,d: \exists \lambda_{i} \neq 0 \}$ satisfying
\begin{equation} \label{eq:contradiction_first_precoder}
\lambda_{1}\mathbf{v}_{1,1} + \lambda_{2}\mathbf{v}_{1,2} + \cdots + \lambda_{d}\mathbf{v}_{1,d} = \mathbf{0},
\end{equation}
where $\mathbf{v}_{i,j}$ is the $j$-th column in $\mathbf{V}_{i}$.

The proof consists of two basic steps. First, we show that the assumption of \eqref{eq:contradiction_first_precoder} leads to one of two consequences: either $\mathcal{N}\big( \mathbf{C}_{2} \big)$ has non-zero vectors, where $\mathbf{C}_{2}$ consists of the last $(K_{t}N_{t} - N_{t})$ columns in $\mathbf{C}$, or $\tilde{\mathbf{V}}$ has linearly dependent vectors, where $\tilde{\mathbf{V}} = \big[ \mathbf{V}_{1}^{T}, \mathbf{V}_{2}^{T}, \cdots, \mathbf{V}_{K_{t}}^{T} \big]^{T}$ is a solution to \eqref{eq:design_precoders_for_users_in_one_cell_extended}. Then, we show that these two consequences are impossible; thus, \eqref{eq:contradiction_first_precoder} cannot occur, and hence $\mathbf{V}_{1}$ must be a full column rank matrix.

The details of the proof are given as follows. As $\tilde{\mathbf{V}}$ is the solution to \eqref{eq:design_precoders_for_users_in_one_cell_extended}, we have
\[
\mathbf{C}\tilde{\mathbf{v}}_{i} = \mathbf{0},\ i=1,2,\cdots,d,
\]
where $\tilde{\mathbf{v}}_{i}$ is the $i$-th column in $\tilde{\mathbf{V}}$. These equations lead to
\[
\mathbf{C} \lambda_{i} \tilde{\mathbf{v}}_{i} = \mathbf{0},\ i=1,2,\cdots,d,
\]
and hence,
\begin{equation} \label{eq:column_sol}
\mathbf{C} \sum_{i=1}^{d} \lambda_{i} \tilde{\mathbf{v}}_{i} = \mathbf{0}.
\end{equation}
As $\mathbf{V}_{1}$ consists of the first $N_{t}$ rows in $\tilde{\mathbf{V}}$, \eqref{eq:contradiction_first_precoder} is equivalent to
\begin{equation} \label{eq:v_tilde_first_part}
\big( \sum_{i=1}^{d} \lambda_{i} \tilde{\mathbf{v}}_{i} \big)_{1:N_{t}} = \mathbf{0},
\end{equation}
where $\big( \sum_{i=1}^{d} \lambda_{i} \tilde{\mathbf{v}}_{i} \big)_{1:N_{t}}$ consists of the first $N_{t}$ elements in $\sum_{i=1}^{d} \lambda_{i} \tilde{\mathbf{v}}_{i}$. Therefore, \eqref{eq:column_sol} can be rewritten as
\begin{equation} \label{eq:column_sol_2}
\begin{bmatrix}
\mathbf{C}_{1} & \mathbf{C}_{2}
\end{bmatrix} \begin{bmatrix} \mathbf{0} \\ \big( \sum_{i=1}^{d} \lambda_{i} \tilde{\mathbf{v}}_{i} \big)_{(N_{t} + 1):K_{t}N_{t}} \end{bmatrix} = \mathbf{0},
\end{equation}
where $\mathbf{C}_{1}$ and $\big( \sum_{i=1}^{d} \lambda_{i} \tilde{\mathbf{v}}_{i} \big)_{(N_{t} + 1):K_{t}N_{t}}$ denote the matrix consisting of the first $N_{t}$ columns in $\mathbf{C}$ and the vector consisting of the last $(K_{t}N_{t} - N_{t})$ elements in the vector $\sum_{i=1}^{d} \lambda_{i} \tilde{\mathbf{v}}_{i}$, respectively. From \eqref{eq:column_sol_2}, we can deduce that
\begin{equation} \label{eq:column_sol_3}
\mathbf{C}_{2} \big( \sum_{i=1}^{d} \lambda_{i} \tilde{\mathbf{v}}_{i} \big)_{(N_{t} + 1):K_{t}N_{t}} = \mathbf{0}.
\end{equation}
As $\mathbf{C}_{2}$ is a full column rank matrix according to \textit{Lemma \ref{lemma:rank_coef_matrix}} and \eqref{eq:constraint_kappa}, $\mathcal{N}\big( \mathbf{C}_{2} \big)$ cannot have non-zero vectors, i.e., the first consequence cannot be the case; thus, \eqref{eq:column_sol_3} is possible only if
\begin{equation} \label{eq:v_tilde_last_part}
\big( \sum_{i=1}^{d} \lambda_{i} \tilde{\mathbf{v}}_{i} \big)_{(N_{t} + 1):K_{t}N_{t}} = \mathbf{0}.
\end{equation}
By combining \eqref{eq:v_tilde_last_part} and \eqref{eq:v_tilde_first_part}, we have
\[
\sum_{i=1}^{d} \lambda_{i} \tilde{\mathbf{v}}_{i} = \mathbf{0},
\]
which means that $\tilde{\mathbf{V}}$ has linearly dependent vectors. However, as $\text{dim}\big( \mathcal{N}(\mathbf{C}) \big) \geq d$, $\tilde{\mathbf{V}}$ can always be chosen to have linearly independent vectors, i.e., $\tilde{\mathbf{V}}$ can always be chosen so that the second consequence cannot be the case. Therefore, both consequences are impossible, and hence $\mathbf{V}_{1}$ is of full column rank.

The same argument can also be used to show that $\mathbf{V}_{i},\ i=2,3,\cdots,K_{t}$ are of full column rank.

\bibliographystyle{IEEEtran}
\bibliography{IEEEabrv,references}

\end{document}